\documentclass[12pt,preprint]{aastex}

\shorttitle{NH$_3$(6,6) in IC342}
\shortauthors{Montero-Casta{\~n}o et al.}

\begin{document}

\title{HOT MOLECULAR GAS IN THE NUCLEAR REGION OF IC~342}

\author{Mar{\'{\i}}a Montero-Casta{\~n}o\altaffilmark{1,2}, Robeson M. Herrnstein\altaffilmark{3} and Paul T.P. Ho\altaffilmark{1,4}}
\altaffiltext{1}{Harvard-Smithsonian Center for Astrophysics, 60 Garden St., Cambridge, MA 02138, mmontero@cfa.harvard.edu, pho@cfa.harvard.edu}
\altaffiltext{2}{Departamento de Astrof{\'{\i}}sica, Facultad de Ciencias F{\'{\i}}sicas, Universidad Complutense de Madrid, 28040-Madrid, Spain}
\altaffiltext{3}{Department of Astronomy, Columbia University, 550 West 120th St., New York, NY 10027, herrnstein@astro.columbia.edu} 
\altaffiltext{4}{Academia Sinica Institute of Astronomy and Astrophysics, Taipei, Taiwan}
\altaffiltext{5}{The National Radio Astronomy Observatory is a facility of the National Science Foundation operated under cooperative agreement by Associated Universities, Inc.} 
\altaffiltext{6}{This research has made use of the NASA/IPAC Extragalactic Database (NED) which is operated by the Jet Propulsion Laboratory, California Institute of Technology, under contract with the National Aeronautics and Space Administration}

%%%%%%%%%%%%%%%%%%%%%%%%%%%%%%%%% ABSTRACT %%%%%%%%%%%%%%%%%%%%%%%%%%%%%%%%%%

\begin{abstract}
We present the first interferometric detection of extragalactic
NH$_3$(6,6) emission in the nearby galaxy IC~342 made using the
VLA$^{5}$.  The data have a resolution of 7.8''~$\times$~5.0'' and
trace hot (T $\sim$ 412 K) and dense ($>$~10$^{4}$~cm$^{-3}$)
molecular gas. We have covered a 170''$\times$300'' area, and detect
two very strong line emission peaks, likely associated with the two
strongest star formation regions of the central part of the
galaxy. We compare these emission peaks to CO (1-0) and (2-1) emission data, which are the most abundant CO
transitions and trace spatially extended emission.  The NH$_3$(6,6) emission is
also compared to emission data from three high-density, nitrogen bearing tracers:
HNC(1-0), HC$_3$N(10-9) and N$_2$H$^+$. Our results suggest that the
molecular mass in the nuclear region of IC~342 has at least two
different components, a dense and cold component and a less dense and
hotter component.

\end{abstract}

%%%%%%%%%%%%%%%%%%%%%%%%%%%%%%% KEYWORDS %%%%%%%%%%%%%%%%%%%%%%%%%%%%%%%%%%%%


%%%%%%%%%%%%%%%%%%%%%%%%%%%%%% INTRODUCTION %%%%%%%%%%%%%%%%%%%%%%%%%%%%%%%%%

\section{Introduction}
IC~342, a late-type spiral galaxy (Scd), is one of the closest
galaxies to the Milky Way, located only $\sim$~3.3~Mpc away
\citep{sah02,kar05} (1'' corresponds to 16~pc). The physical
properties of its molecular clouds, its infrared luminosity, and the
presence of a nuclear stellar cluster \citep{hut95,schu01} make IC~342 similar
to the Milky Way in many ways. Because of its proximity to the
Galactic equator, IC~342 is a faint optical source.  However, it is a
very rich source of molecular emission, and has been studied in numerous
molecular tracers. Furthermore, IC~342 is nearly face-on, minimizing
line broadening due to galactic rotation, thereby maximizing line
intensity.  This orientation makes IC~342 useful for Milky Way-like
studies.

IC~342 has a high concentration of molecular gas within the inner
1~kpc, especially within 250~pc of the nucleus \citep{isr03}. Inflow
along a large-scale stellar bar \citep{schi03} has been proposed to
explain the large quantity of gas in the nucleus. An S-shaped bar in
the molecular gas (the ``mini-spiral''), has been clearly detected in
the nuclear region in CO(1-0) \citep{hel03} as well as in NIR
observations \citep{bok97}. The mini-spiral is oriented in the
north-south direction, and it is apparently unrelated to the spiral
arms located further out in the disk \citep{lo84}. The two arms of the
mini-spiral meet in a ring composed of dense gas
\citep{ish90,mei01}. This ring, $\sim$ 4'' in radius, surrounds a
central stellar cluster that occupies the central 80~pc of the galaxy
\citep{tur92,bok97}. The central stellar cluster is non-axisymmetric,
with a north-south elongation, which is believed to be caused by the
presence of a nuclear stellar bar with a major axis diameter of $\sim$~13''\citep{bok97}. The nuclear bar creates the ring-like structure at 4'', an Inner Lindblad Resonance (ILR) \citep{bok97}.  Inside the ring, the
abundance decreases for all the molecular tracers studied to date
\citep{mei05}.

Current star formation activity in the nucleus of IC~342 is dominated
by two bright HII regions $\sim$ 4''west and east of the nuclear
cluster. These star formation regions are $\sim$~5~Myr old.  The
central cluster, which is composed of young and massive supergiants,
does not show evidence for current star formation \citep{bok97}.

High-resolution HCN(1-0) studies by \citet{dow92} have shown the
presence of five giant molecular clouds (GMCs) in IC~342 (GMCs A, B,
C, D, and E). These GMCs have sizes from 20 to 40 pc, masses of
$\sim$$10^{6}$ M$_\odot$ and average molecular densities of
$\sim10^4$ cm$^{-3}$ \citep{schu01}. Each GMC appears to be composed of multiple
small, dense clouds \citep{mei01}, which have densities of
$\sim$~10$^{6}$~cm$^{-3}$ \citep{schu01}). \citet{dow92},
\citet{hel93} and \citet{jac96} have shown that the distribution of
GMCs extends close to the nucleus (within a distance of 100~pc) and
well inside the tidal limit. The region occupied by these GMCs
includes both the arms of the mini-spiral and the ring.
 
GMCs A, B and C are located in the central ring, whereas GMCs D and E
are in the northern arm and southern arm of the mini-spiral,
respectively. GMCs B and C are situated where the arms of the
mini-spiral meet the ring, and may be related to the young star
formation regions found there (the two HII regions mentioned above).
High-resolution studies of C$^{18}$O by \citet{mei01} show that GMC B
is the warmest GMC in the central region. GMC A is the closest to the
nucleus, but it does not show strong star formation. GMC D also does
not show strong star formation, although it is located alongside the
northern arm where \citet{tur92} detected large quantities of
molecular gas with inward motions. GMC N, as noted by \citet{mei05},
is located close to GMCs A and B, in the molecular ring. It has only
been detected in nitrogen bearing molecules.

The GMCs are very warm, with a temperature of 50 to 70~K measured
using HCN(1-0) (\citet{dow92}). However, more recent CO(2-1)
observations by \citet{mei00} have shown that most of the gas within
the GMCs has a temperature of only 10-20~K.  This cool gas is
surrounded by much hotter Photon Dominated Regions (PDRs) with
temperatures that reach 50~K. The ionization of the PDRs may be
produced by the central nuclear cluster, rather than by the HII
regions close to GMCs B and C \citep{mei05}.

The GMCs associated with the nucleus of IC~342 have been the subject
of numerous molecular line studies. However, few observations have been made in the rotation inversion transitions of NH$_3$.  NH$_3$ is a good probe
of the physical conditions in molecular clouds for many reasons.  The
large number of transitions concentrated around $\sim$1.3~cm means
that multiple ammonia lines can be detected with the same receiver on
the same telescope, thereby circumventing cross calibration
problems. Moreover, measuring two emission lines allows us to
calculate the rotational temperature of the gas \citep{ho83}. Also,
due to its high dipole moment, a density of $\sim$ 10$^4$ cm$^{-3}$ is
necessary to excite NH$_3$, making it an excellent high-density gas
tracer.

The first detection of extragalactic NH$_3$ emission was made in IC
342 by \citet{mar79} using the 100m Effelsberg telescope of the
Max-Plank-Institut f{\"u}r Radioastronomie. Numerous studies, both
with interferometers and single-dish antennas, have been carried out
since then in IC~342 \citep{ho82,ho83,mar86,ho90b,mau03}, as well as
in other galaxies \citep{hen00,wei01,tak02,beu05,ott05}. However, the
only previous study of NH$_3$(6,6) in IC~342 was made by
\citet{mau03} using the 100m Effelsberg telescope. This study
detected the presence of warm gas ($>$~150~K) in the center of IC~342.
In particular, they found a high abundance of NH$_3$(6,6)
(N$_{66}$~=~3.2~$\times$~10$^{12}$~cm$^{-2}$), which has an excitation energy
above ground of 412~K. This study also suggested the presence of
multiple components with different temperatures within the
beam. However, due to the large beam (38'' for NH$_3$(6,6)) for these
single-dish data, the multiple components could not be separated.

We have conducted a study of NH$_3$(6,6) using the VLA in order to
resolve the structure of the hot dense gas in the nucleus of
IC~342. Our interferometric observations produce a much smaller beam
(7.8''~$\times$~ 5.0''), allowing us to separate the emission from
different structures present in the nucleus of IC~342.  In addition to
studying the molecular structure of the IC~342 nuclear region, we also
relate our data to the center of the Milky Way.  As previously
mentioned, IC~342 is a perfect candidate for Milky Way-like studies.
Interferometric observations of NH$_3$(6,6) at the Galactic Center
have shown that there is a large concentration of high density hot gas
surrounding Sgr~A* \citep{her02,her05}. Therefore, the study of the nuclear
region of IC~342 is not only important in itself, but it also provides
an excellent opportunity to investigate the same processes that take
place in our own Galactic Center.

%%%%%%%%%%%%%%%%%%%%%%%%%%%%%%% OBSERVATIONS %%%%%%%%%%%%%%%%%%%%%%%%%%%%%%%%

\section{Observations}

Observations of NH$_3$(6,6) ($\nu$~=~25.056025~GHz) were made with the
NRAO Very Large Array (VLA$^{5}$) in the D configuration on 2003
February 15 and 16. The pointing was centered on
$\alpha_{2000}$~=~03$^h$46$^m$48.7$^s$,
$\delta_{2000}$~=~68\arcdeg05'46.7''.  The data consisted of a 25 MHz
bandwidth divided into 15 spectral channels with a velocity resolution
of 9.3~km~s$^{-1}$. The velocity coverage was from -30~km~s$^{-1}$ to
+110~km~s$^{-1}$, centered on {\it
v}$_{LSR}$~=~40~km~s$^{-1}$. Channels 1, 14 and 15 were removed from
the data because of low sensitivity due to the roll off in the
passband response. Channels 2 and 3 were used for continuum
subtraction, which was performed in the uv plane, thus resulting in an
effective velocity coverage from 0~km~s$^{-1}$ to 90~km~s$^{-1}$. The
fact that the continuum subtraction has been performed using channels
from only one side of the line does not affect the final result since
the bandpass was calibrated by observing 0319+415 (3C84).

The data were calibrated using the NRAO Astronomical Imaging
Processing System (AIPS). The total on-source integration time was
$\sim$7~hours. The flux calibrator was 0137+331 (3C48) with an assumed
flux of 0.99 Jy at 1.3~cm. The phase calibrator was 0228+673, with a measured
flux of 1.04~$\pm$~0.01~Jy. We performed phase-only self-calibration
to correct for atmospheric fluctuations on short time-scales. The
self-calibration was done on the broadband continuum data (the average
of the inner 75\% of the 25 MHz window), and the results were applied
to the line data. Final deconvolution was performed using IMAGR. The
overall improvement in RMS noise due to self-calibration was about a
factor of 2. The final RMS noise per channel is 0.3~mJy~Beam$^{-1}$
and the overall achieved RMS sensitivity is
3.6~mJy~Beam$^{-1}$~km~s$^{-1}$. Natural weighting of the uv data
produced an image with a synthesized beam of 7.8'' x 5.0'' with a
position angle of 64\arcdeg. The integrated intensity map was produced
using the MOMNT task with a minimum flux cutoff of 0.4~mJy.  The
absolute positional accuracy is limited by phase noise, and is on the
order of 1''.

%%%%%%%%%%%%%%%%%%%%%%%%%%%%%%%%%% RESULTS %%%%%%%%%%%%%%%%%%%%%%%%%%%%%%%%%% 
\section{Results}

\begin{figure}
\begin{center}
\epsscale{1}
\plotone{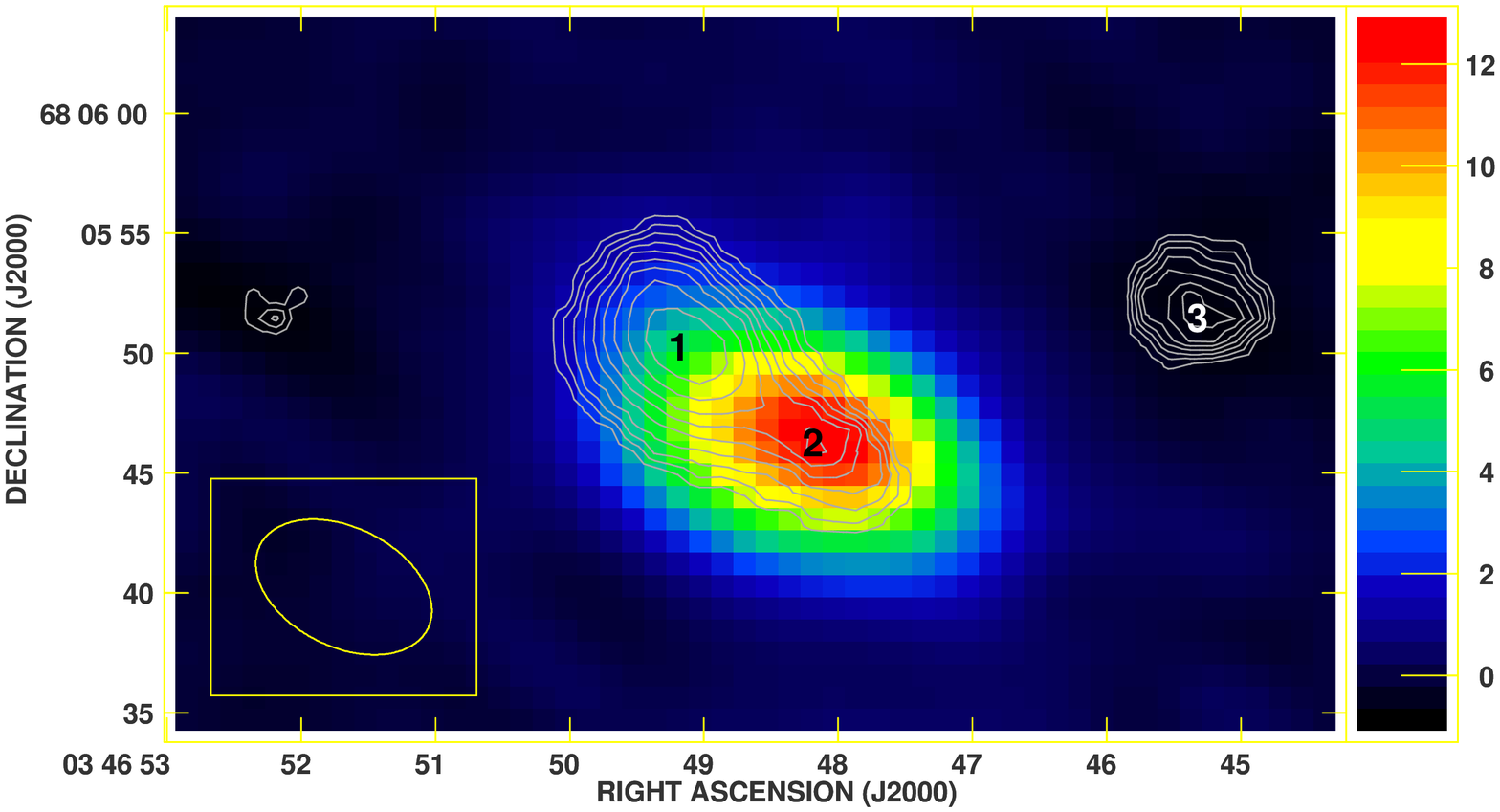}
\end{center}
\caption{NH$_3$(6,6) integrated intensity map. The contours represent the line emission, while the false-color scale indicates the continuum emission at 1.3~cm. Peak 1 represents the {\it (6,6) peak}, peak 2 the {\it continuum peak} and peak 3 the {\it west peak}. The contour levels are in steps of 10\% of the intensity peak, from 5.5~mJy~Beam$^{-1}$~km~s$^{-1}$ to 49.5~mJy~Beam$^{-1}$~km~s$^{-1}$. The RMS is 3.6~mJy~Beam$^{-1}$~km~s$^{-1}$, therefore the contour levels in the map are in steps of $\sim$~1.5~$\sigma$, from 1.5~$\sigma$  to 13.7~$\sigma$. The false-color scale is in mJy~Beam$^{-1}$. \label{cont.fig}}
\end{figure}

We detect and resolve the extended NH$_3$(6,6) emission from IC~342
(figure \ref{cont.fig}). Three line emission peaks are clearly visible
in our results, marked by numbers 1-3 on figure \ref{cont.fig}. Peaks
1 and 2 appear to be embedded in a single, extended molecular
structure. The third peak, which has not been previously observed in
any molecular tracer, is isolated. Comparing the line emission map
with the continuum map (in contours and false colors, respectively, in
figure \ref{cont.fig}), we find that peak 2 corresponds to the
continuum emission peak, which marks the location of a strong heating
source as well as the nucleus of IC~342 \citep{dow92}. Throughout this
paper, peak 1 will be referred to as the {\it (6,6) peak}, peak 2 as
the {\it continuum peak}, and peak 3 as the {\it west peak}.
 
We compared our NH$_3$(6,6) flux with the single-dish data from
\citet{mau03}. Because the beam size for the 100m Effelsberg telescope
is 38'', all three of our NH$_3$(6,6) peaks are well within the
Effelsberg beam. From \citet{mau03}, the single-dish flux is
4.5~$\times$~10$^{-3}$~Jy~Beam$^{-1}$, while the sum of the fluxes
detected for the three NH$_3$(6,6) peaks presented in this paper is
4.6~$\times$~10$^{-3}$~Jy~Beam$^{-1}$. Therefore, there is no missing
flux, and all emission detected in the single-dish data must originate
in clouds with size scales smaller than $\sim8''$.

\begin{figure}
\begin{center}
\epsscale{1}
\plotone{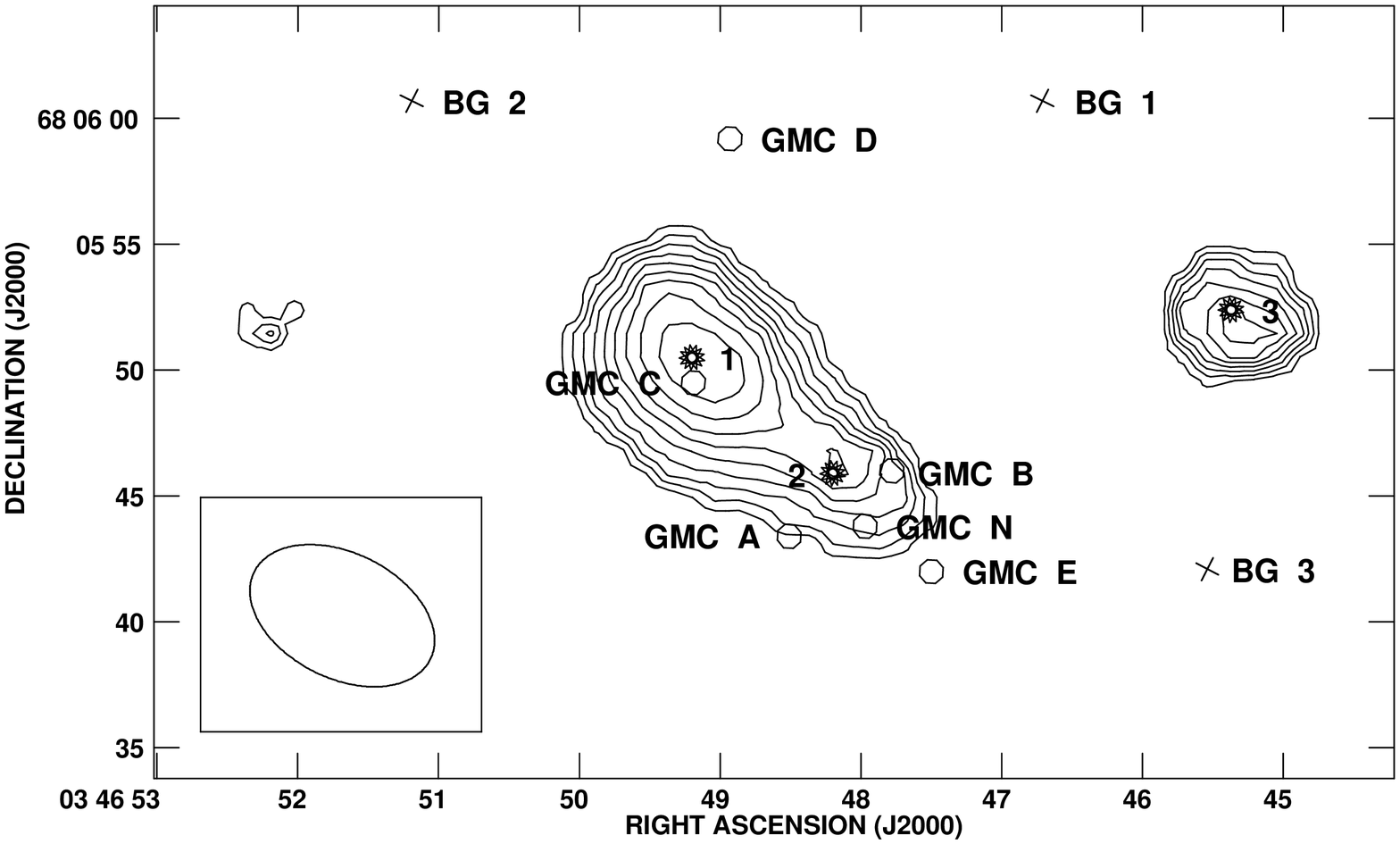}
\end{center}
\caption{NH$_3$(6,6) integrated intensity map. We have marked the positions of the three NH$_3$(6,6) detected peaks as well as those of the 6 GMCs as described by \citet{dow92} and \citet{mei05}. The 6 GMCs are marked with circles, the {\it (6,6) peak} (peak 1), the {\it continuum peak} (peak 2) and the {\it west peak} (peak 3) are marked with stars and numbers. We have also marked three random background positions whose spectra we have plotted (figure \ref{spectra.fig}) for comparison reasons. The background points are marked with crosses and labeled 'BG1', 'BG2' and 'BG3'. \label{positions.fig}}
\end{figure}

\begin{figure}
\begin{center}
\epsscale{0.30}
\plotone{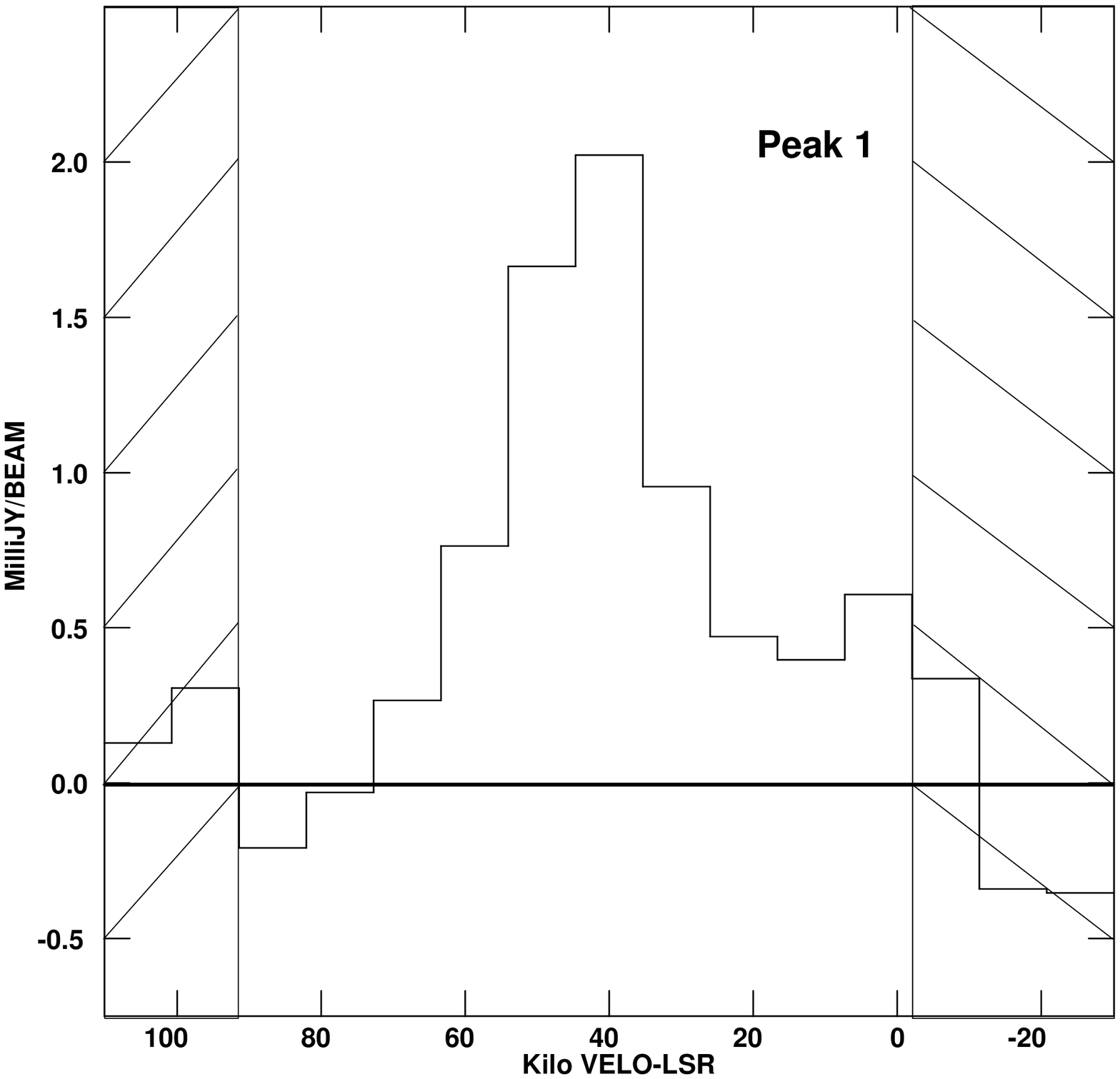}
\plotone{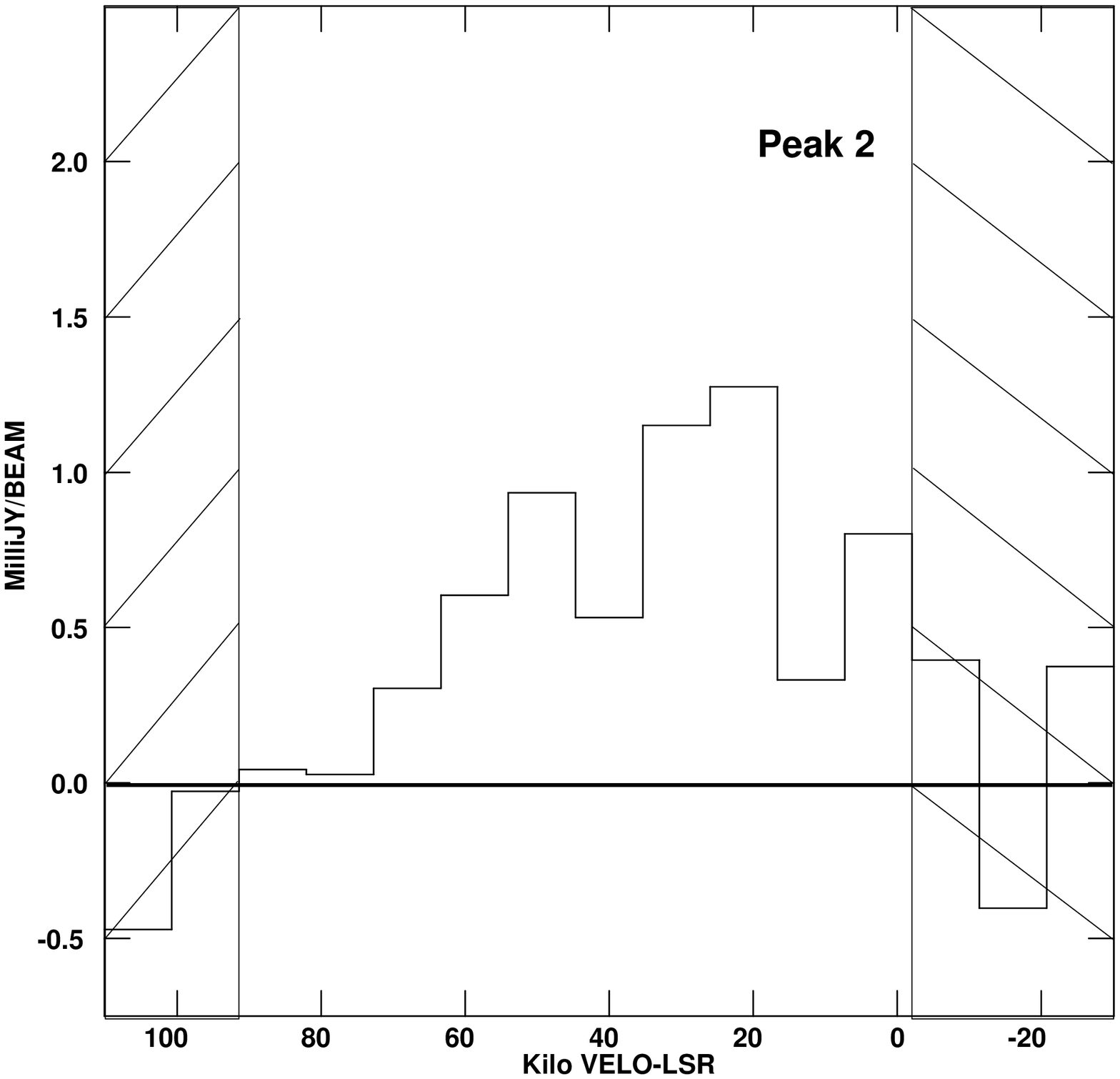}
\plotone{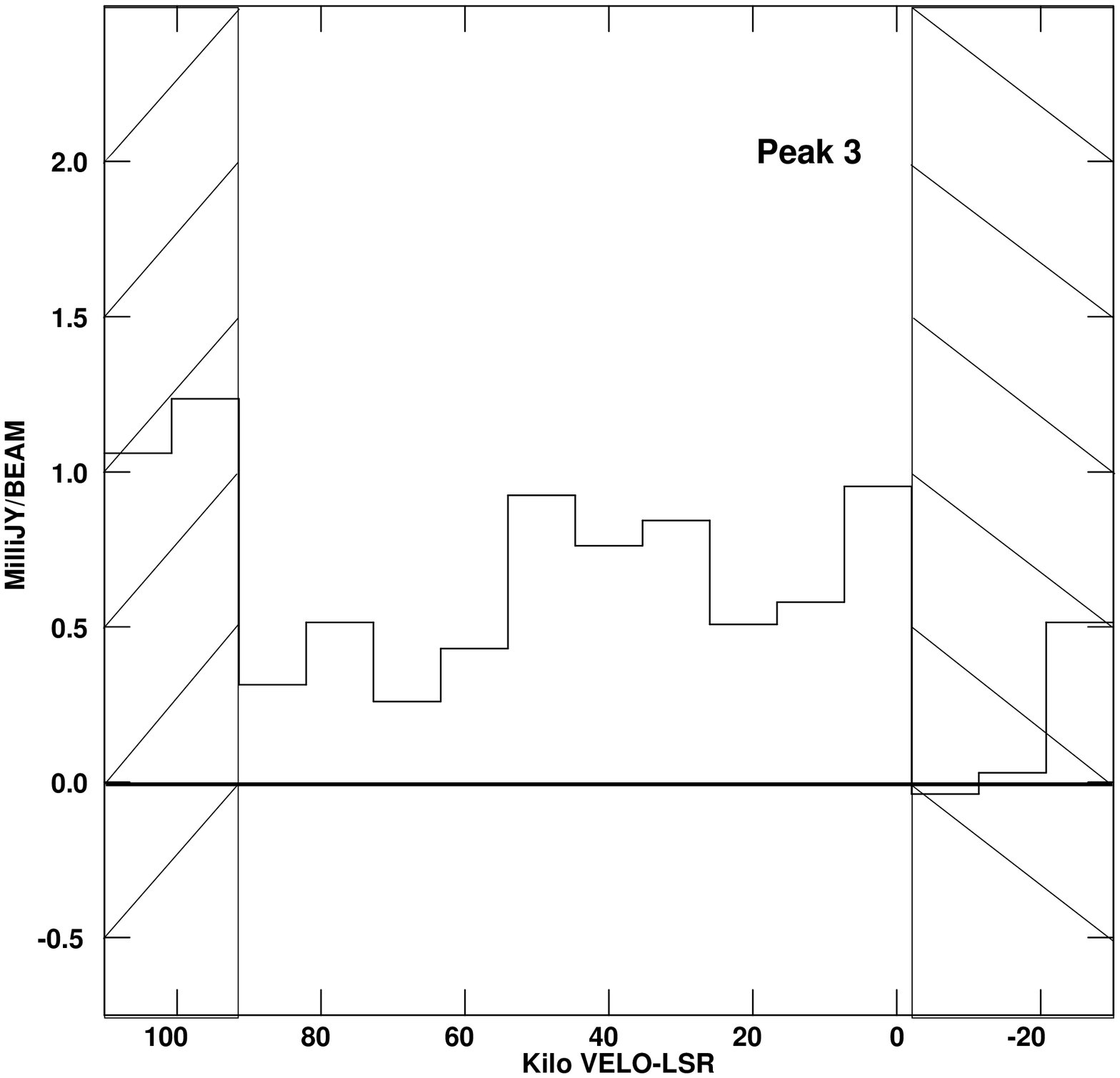}

\plotone{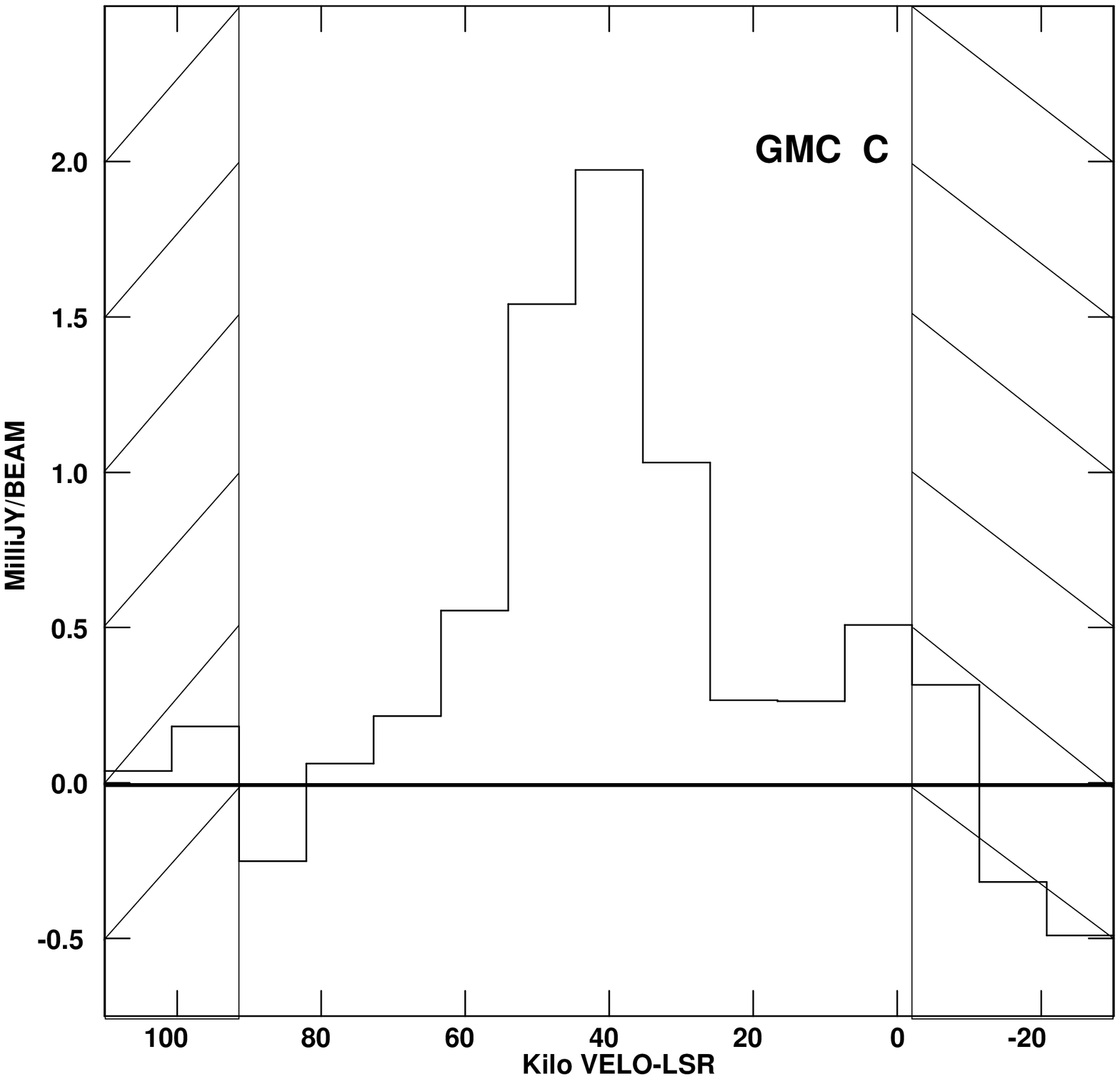}
\plotone{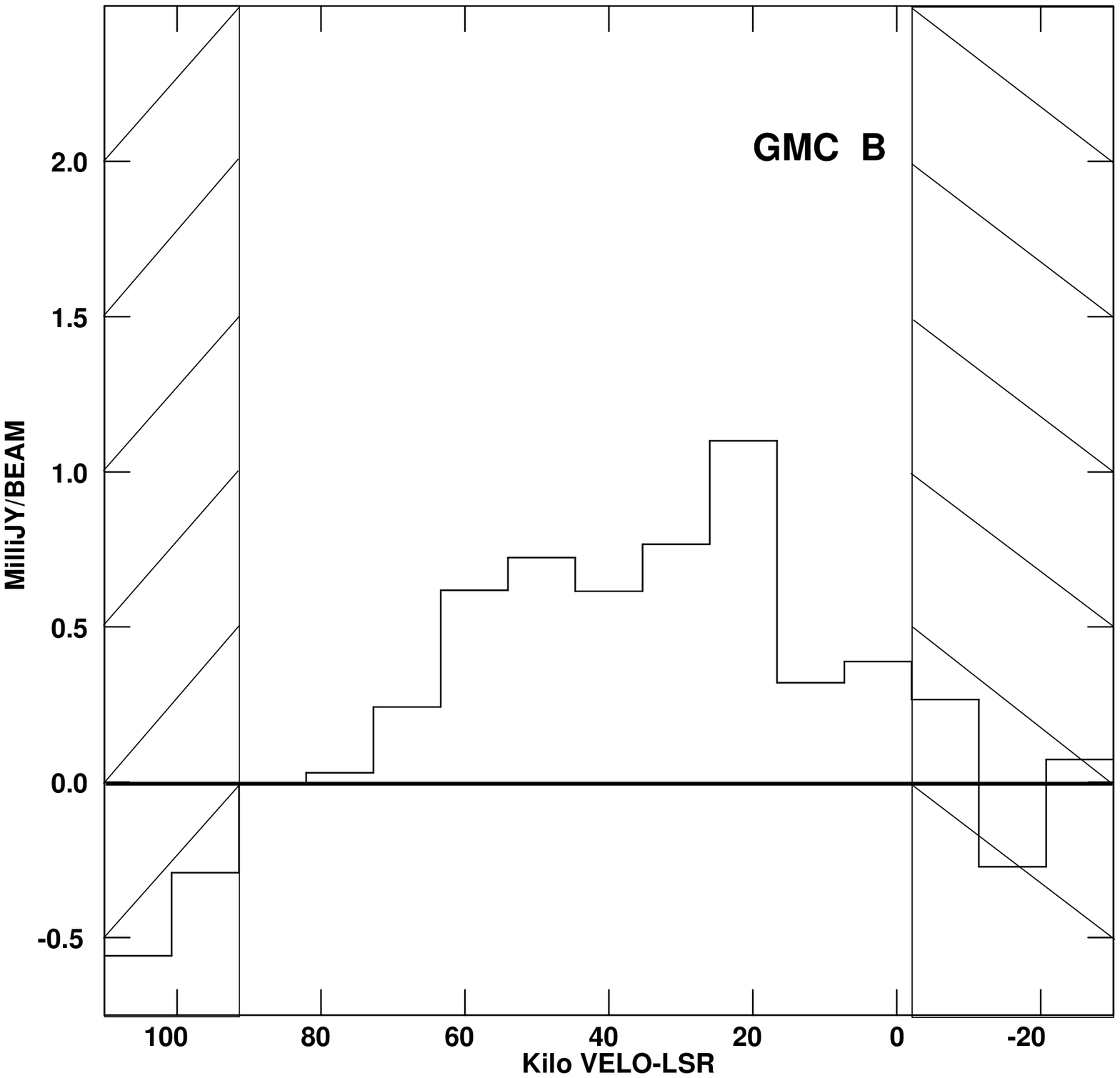}
\plotone{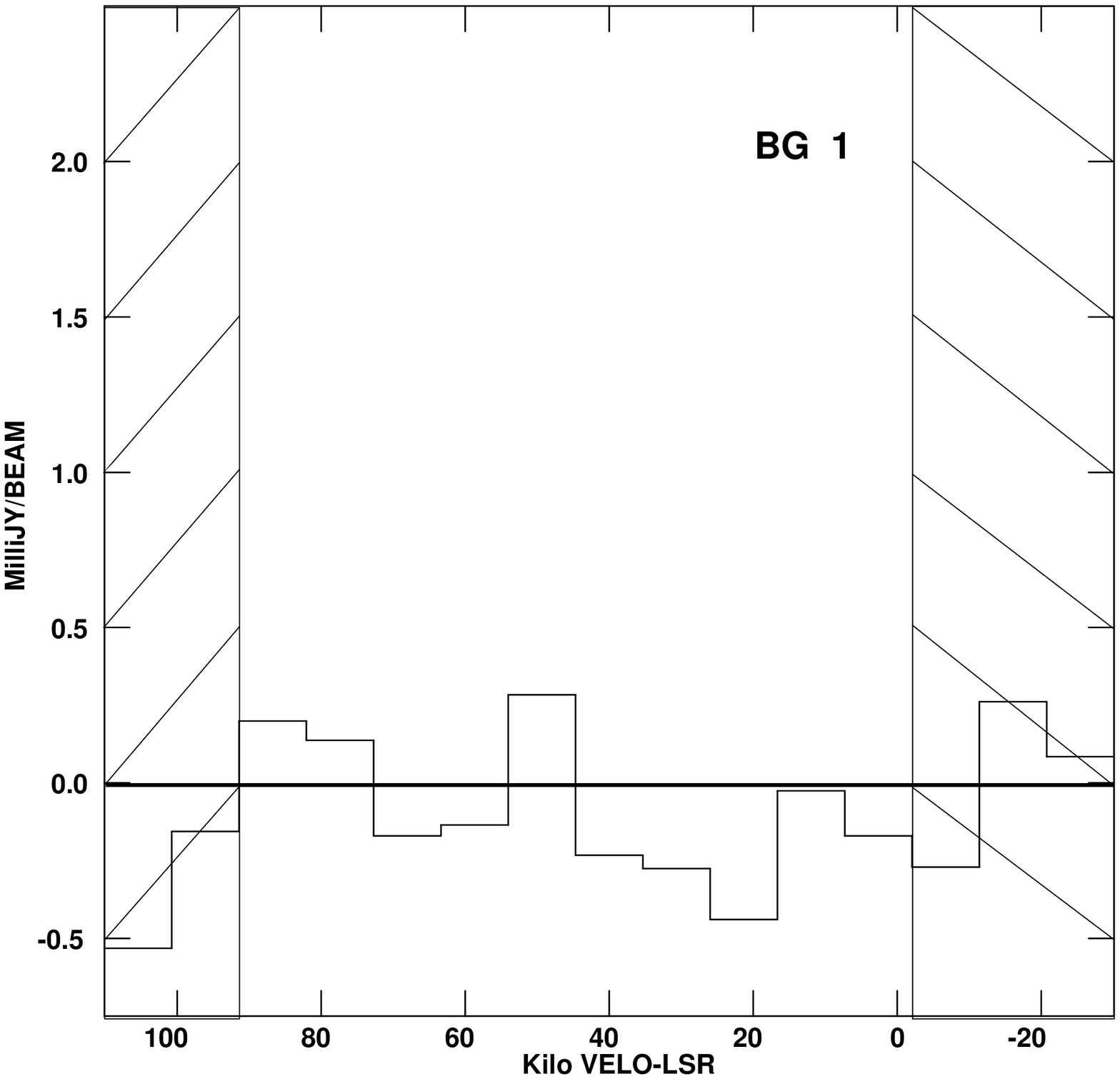}

\plotone{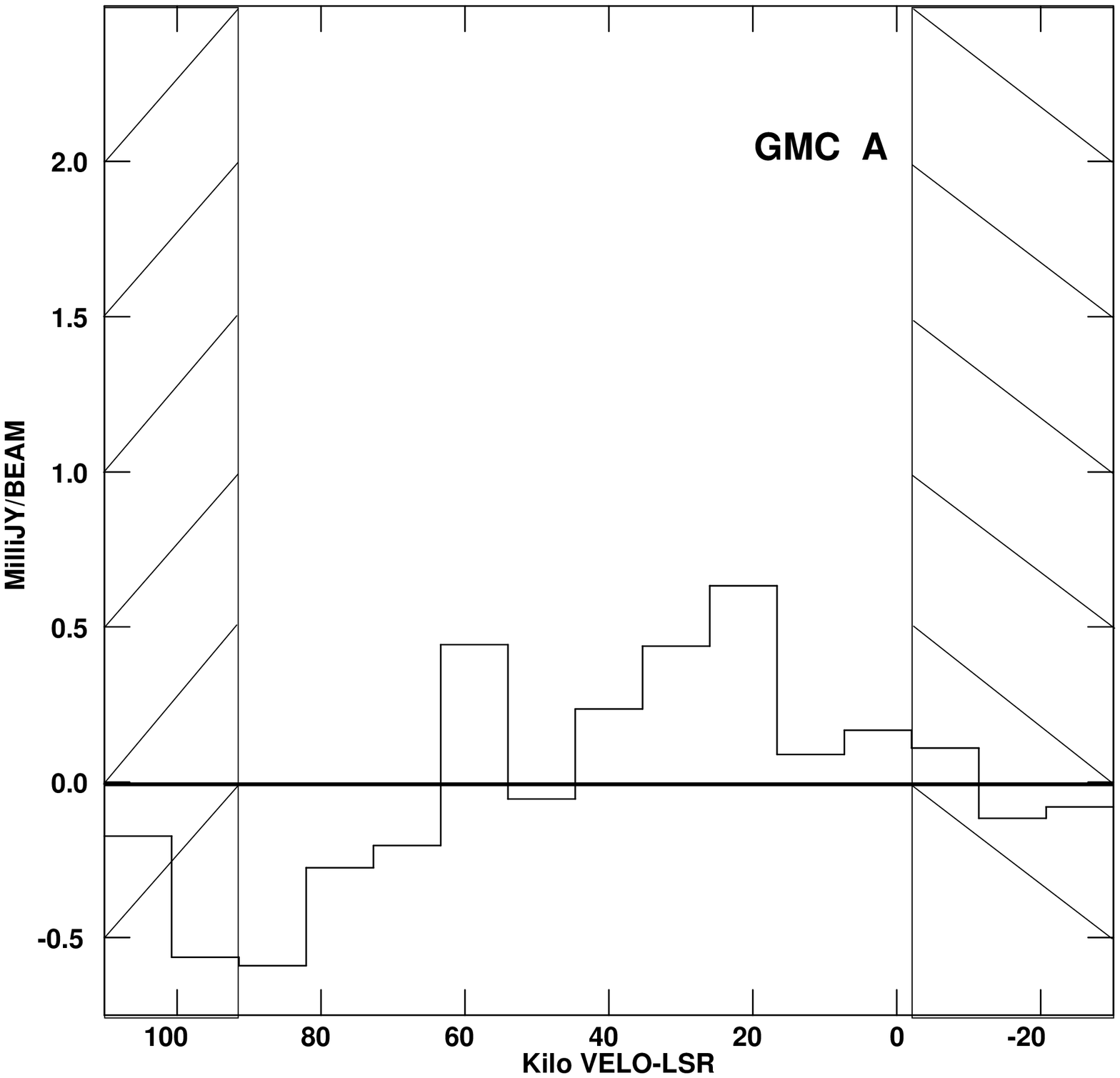}
\plotone{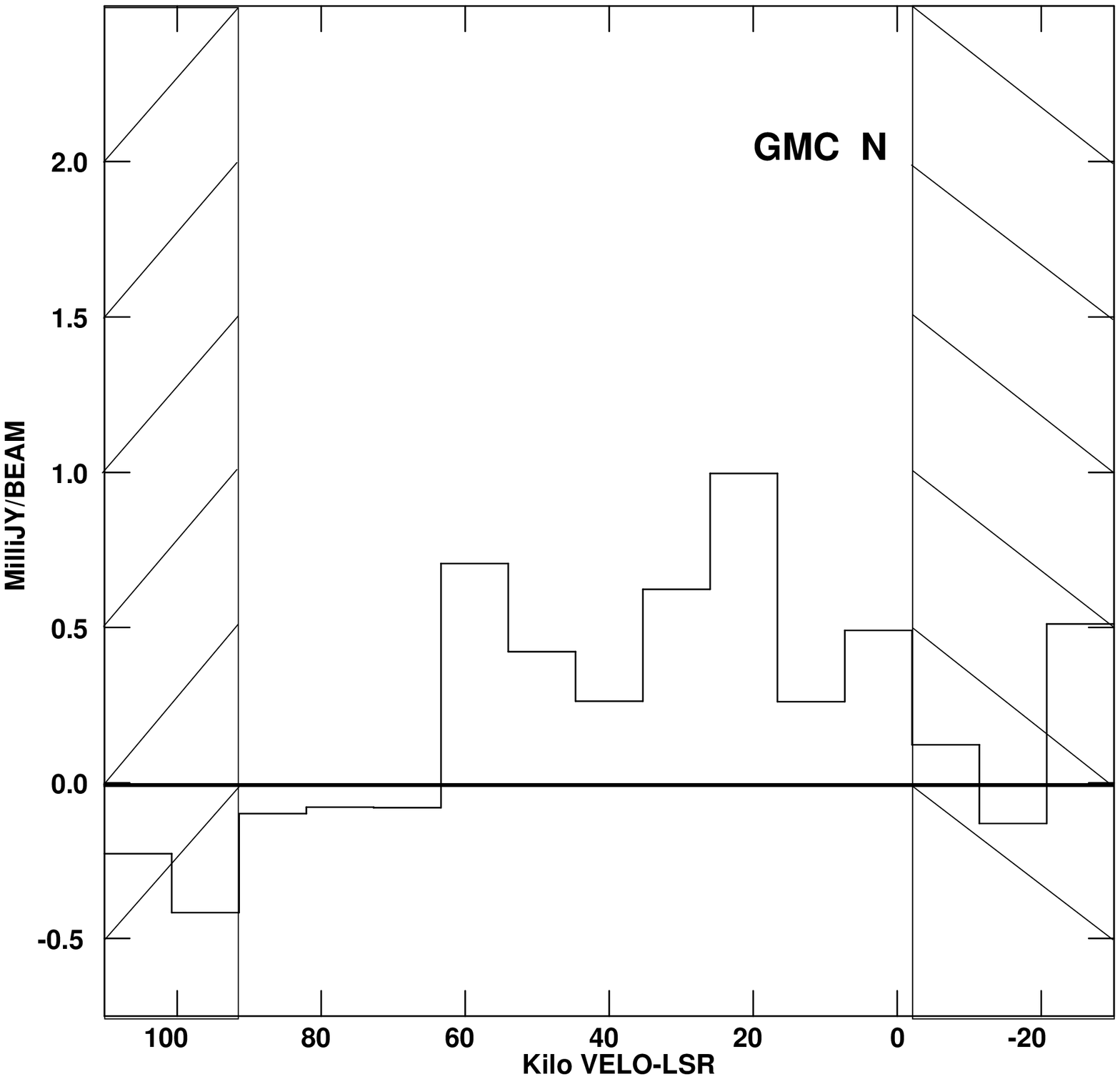}
\plotone{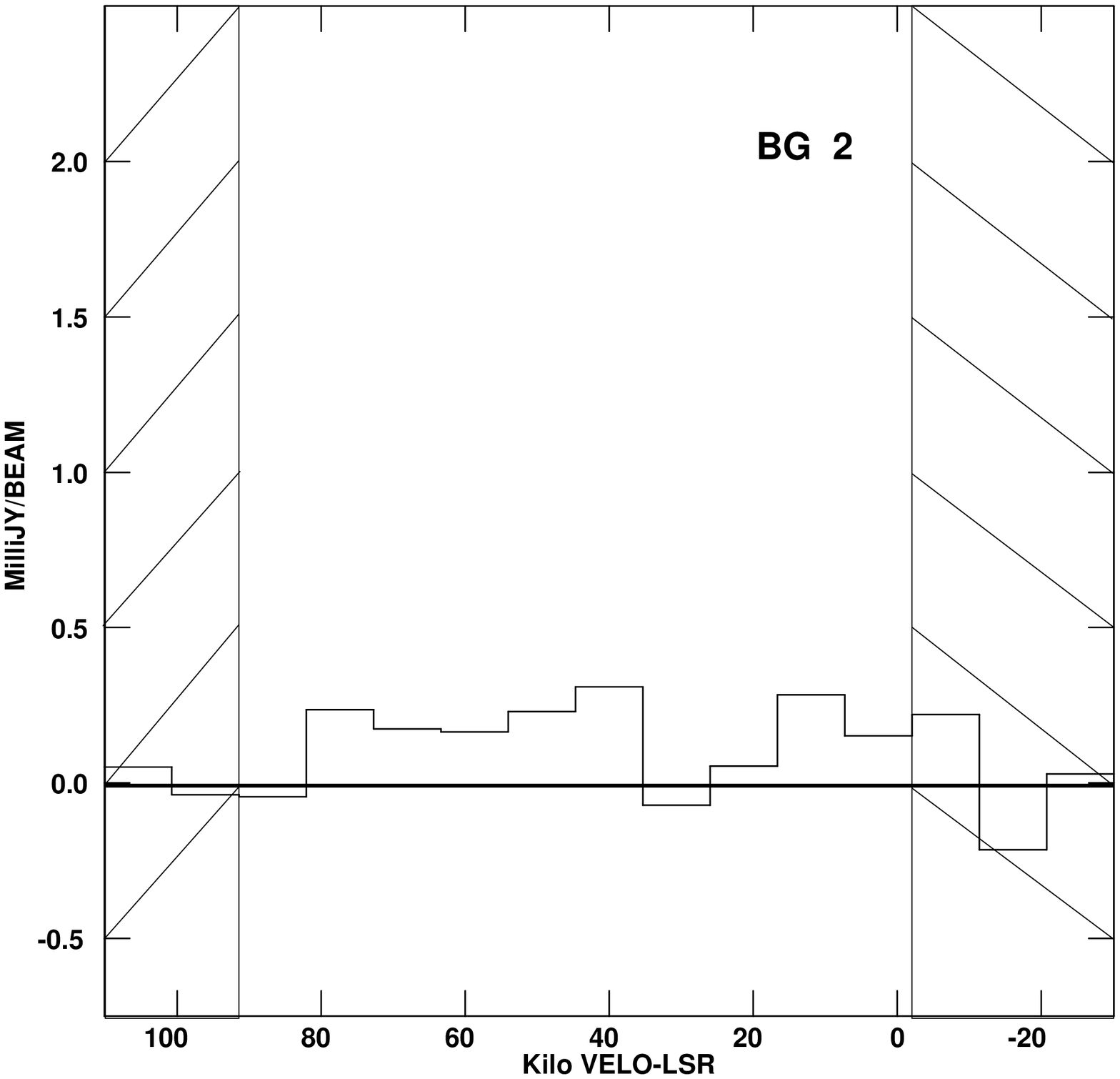}

\plotone{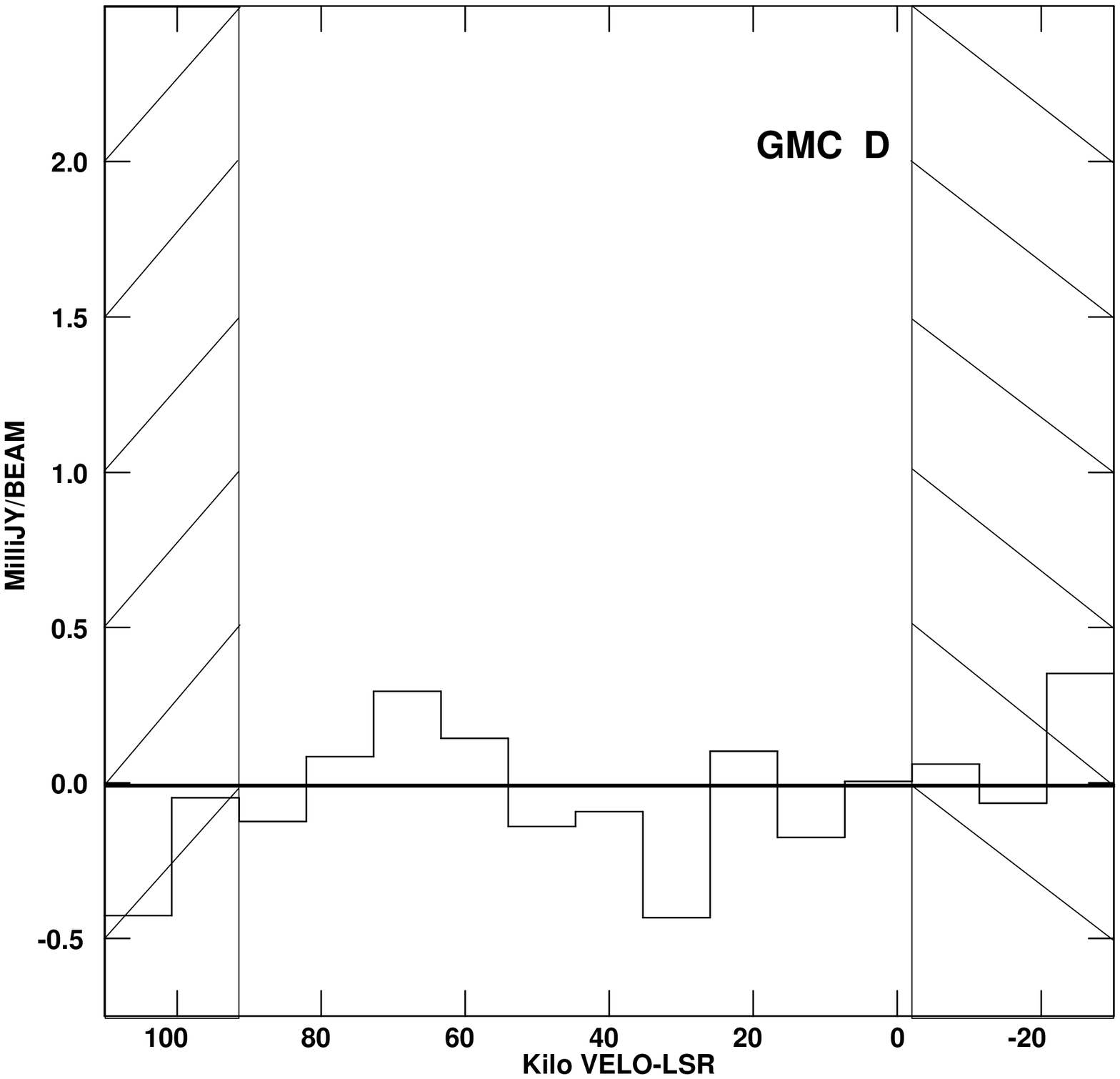}
\plotone{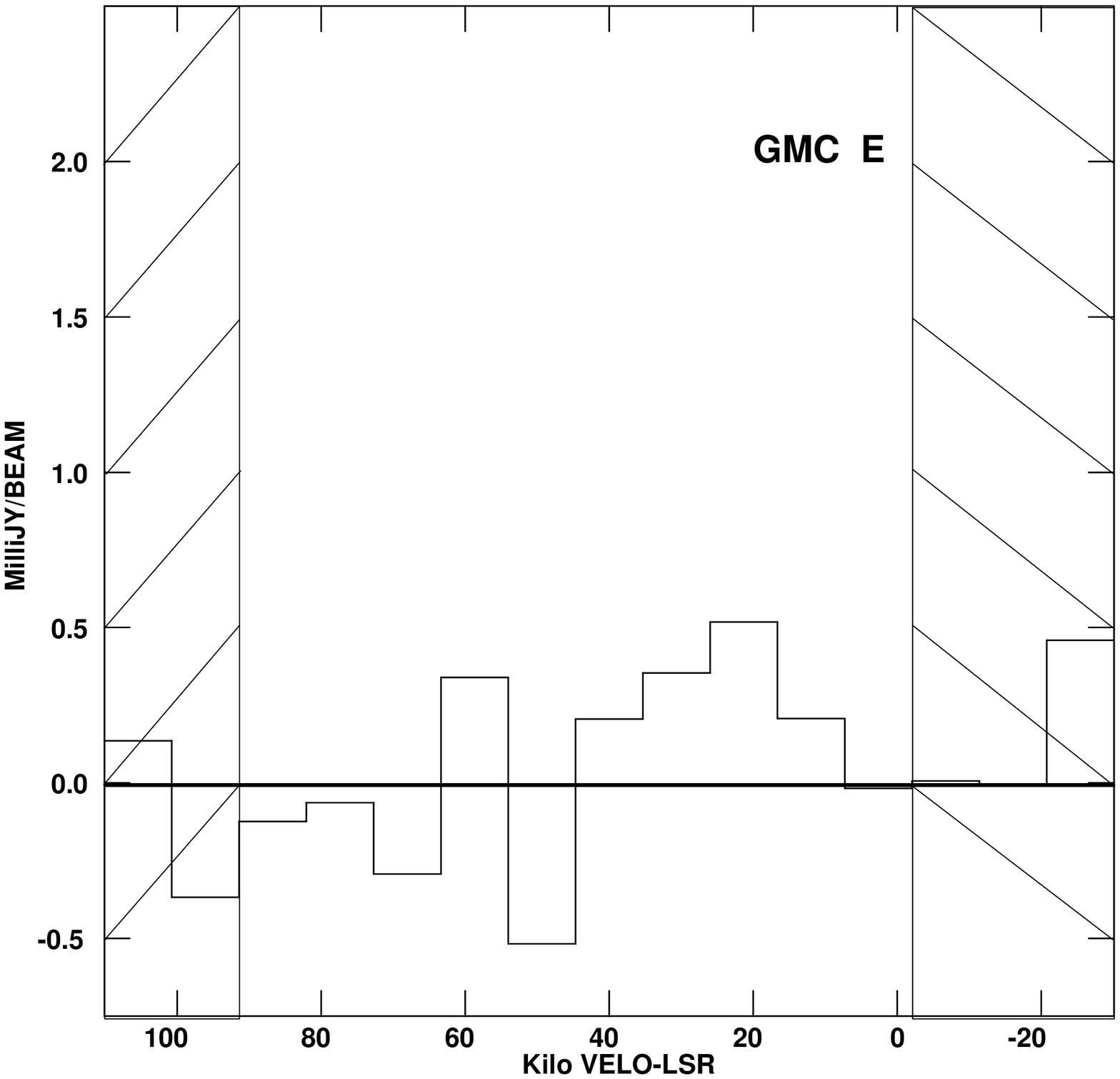}
\plotone{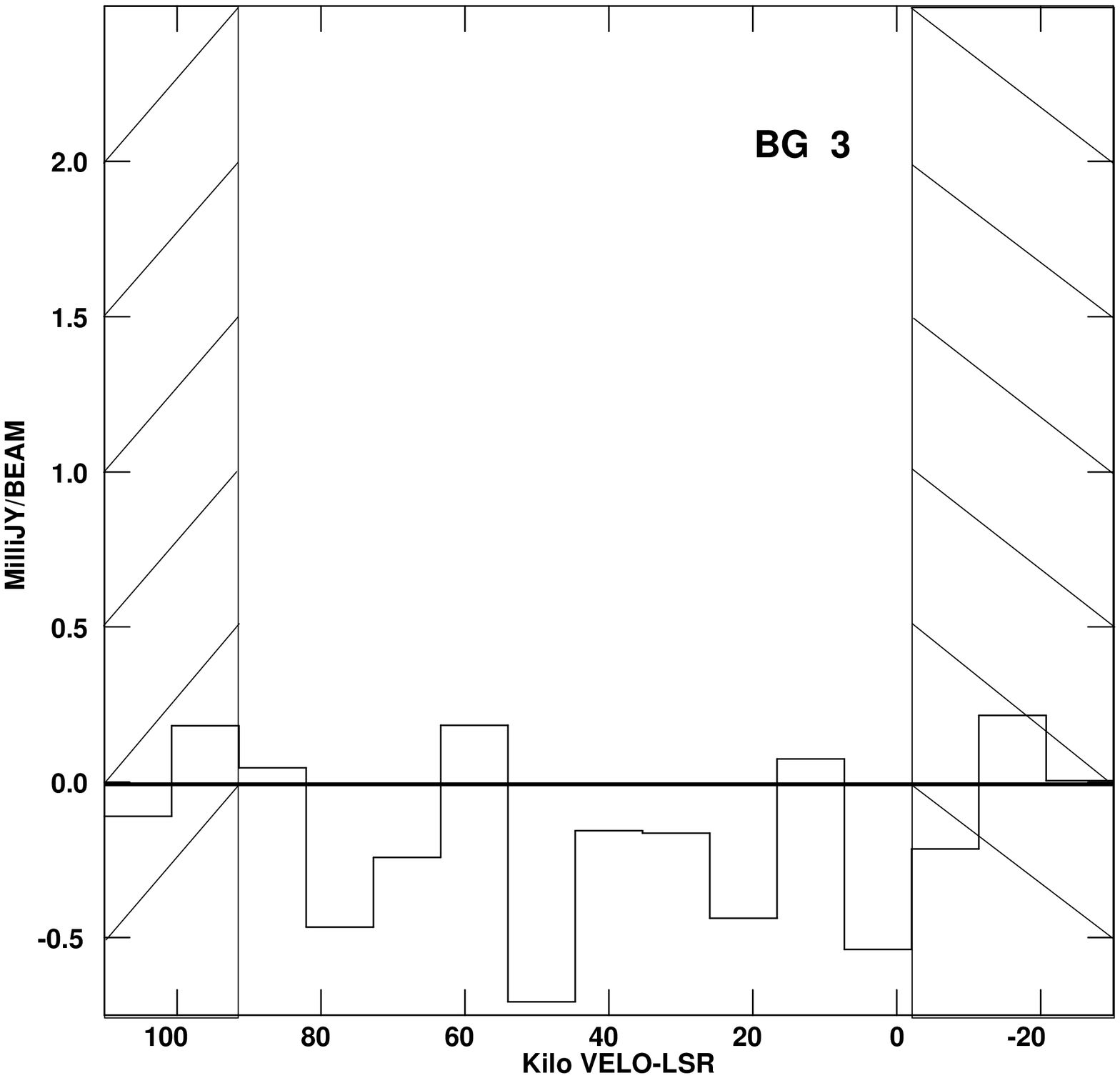}
\end{center}
\caption{Spectra measured at the positions of the 6 GMCs (GMCs A, B, C, D, E and N), the three peaks detected in NH$_3$(6,6) (peaks 1,2 and 3 from figure \ref{positions.fig}) and three random background positions (BG1, BG2 and BG3). The shaded areas represent the channels we have either removed because their noise level is too high (channels 1, 14 and 15), which correspond to velocities from -30 to -20~km~s$^{-1}$ and from -90 to +110~km~s$^{-1}$, or used for continuum subtraction (channels 2 and 3, velocity from -20 to $\sim$0~km~s$^{-1}$).\label{spectra.fig}}
\end{figure}

Because the relative locations of the NH$_3$(6,6) emission peaks and
the GMCs may provide a clue to the behavior of material near the
nucleus of IC~342, we overlay the positions of the 6 GMCs on our
NH$_3$(6,6) map (GMCs A, B, C, D, E and N in figure
\ref{positions.fig}). We plot the spectra at the positions of our
three peaks and the 6 GMCs.  For comparison, we also plot spectra
taken from three random, emission-free positions (figure
\ref{spectra.fig}).

In order to investigate the physical conditions in the NH$_3$(6,6)
clouds that we observe, we also compare our NH$_3$(6,6) map to
HNC(1-0), HC$_3$N(10-9), N$_2$H$^+$(1-0) \citep{mei05}, CO(1-0)
\citep{hel03}, and CO(2-1) \citep{schi03}.  We have chosen these five
molecular lines for different reasons. HNC, HC$_3$N and N$_2$H$^+$
have proved to be high-density tracers. Moreover, these molecules are
nitrogen bearing like NH$_3$. Comparing the results from HC$_3$N, HNC
and N$_2$H$^+$ with NH$_3$, we can determine whether GMC N is also an
emission peak in NH$_3$.

CO is second only to H$_2$ in terms of abundance in the interstellar
medium.  However, H$_2$ cannot be directly studied because of the lack
of a dipole moment. We compare our NH$_3$(6,6) data to CO(2-1) and
(1-0), which, because of their low excitation requirements, are the
molecular tracers that detect more extended emission in IC~342. Most of the CO(1-0)
emission comes from the extended ridges of the mini-spiral, but not
directly from the GMCs embeded in it \citep{dow92}. In contrast,
CO(2-1), at twice the excitation above ground, is detected in warmer
gas than CO(1-0). As mentioned before, \citet{mei00} have found most
of the gas inside the GMCs to be colder than the gas outside of them,
since the GMCs are surrounded by PDRs. Thus, it is not surprising that
the warmer CO(2-1) is detected in the thin outer layers of the GMCs
where they are heated from outside by UV photons
\citep{tur93}. Therefore, the comparison with NH$_3$(6,6) could
provide some clue as to the kind of gas traced by this highly excited
transition. We expect to find NH$_3$(6,6) concentrated in the areas
where CO(1-0) is weaker and inside the GMCs traced by CO(2-1). A gas distribution of this sort would reflect the fact that NH$_3$ is
optically thin, while CO is optically thick, and that the gas
distribution depends on density. Therefore, even though the hotter CO
has been detected outside of the GMCs instead of inside, we expect the
optically thin and dense NH$_3$ to be found well inside the GMCs.

Because the different molecular lines have been measured using
different telescopes, we have resampled the maps with the same size
pixels, in order to allow a proper registration for direct
comparison. We did not convolve the maps, choosing to keep the full
resolution of each map. However, most of the maps we use for
comparison in fact have a very similar angular resolution to our
data. We use the latest HNC(1-0), HC$_3$N(10-9) and N$_2$H$^+$(1-0)
results from \citet{mei05}. These three molecular lines have been
measured using the Owens Valley Radio Observatory (OVRO) and have an
angular resolution of 5-6'' with a positional accuracy of $\sim$
1''. The CO(2-1) map by \citet{schi03} was also made using OVRO, but
the angular resolution is 1.2'' with a positional uncertainty of less
than 0.1''. Matching the angular resolution of CO(2-1) to our NH$_3$
data was not useful because the detected structures would be smoothed
away. CO(1-0) has been measured using the Berkeley-Illinois-Maryland
Array (BIMA) together with the 12m NRAO antenna \citep{hel03}. The
angular resolution of 5.6'' is similar to the resolution of our VLA
data. The CO(1-0) data have a positional accuracy of $\sim$
0.4''. Only the CO(1-0) data include short spacing information, and
are thus sensitive to large-scale structure. Therefore, the flux
associated with the other molecular tracers must be assumed to be a
lower limit. The lack of short spacings does not appear to be as important for NH$_3$(6,6) as there is no missing flux as compared to single-dish measurements. Nevertheless, some caution must be applied to detailed comparisons between molecules.

In order to understand the significance of the {\it west peak}, which
is not detected in any of the other molecular lines, we also compare
our results to a 6~cm continuum map from the VLA (C.-W. Tsai, private
communication) (figure \ref{6cm.fig}). 
%This comparison addresses the role of the synchroton emission associated with the {\it west peak}. 
The 6~cm continuum peak
appears spatially coincident with our 1.3~cm continuum map (figure
\ref{cont_cont.fig}). Therefore, we are confident that the features we
observe in NH$_3$(6,6) are well placed in comparison with the 6~cm
map, i.e. the positional accuracy is reliable.

Finally, it would be very useful to compare the NH$_3$(6,6) emission
to emission from other metastable transitions of
NH$_3$. Comparing our NH$_3$(6,6) map with the NH$_3$(1,1)
and (2,2) maps by \citet{ho90b} we find that, in general, the detected
features on the three maps are remarkably different. Since the lower
excitation lines are sensitive to colder gas, the NH$_3$(1,1) and
(2,2) maps appear to trace the more extended emission, such as the
mini-spiral, and GMC D. However, the emission in these lower lines
seems to ``avoid'' the position where the {\it continuum peak} is
located, whereas the NH$_3$(6,6) emission is very strong in that
location. \citet{ho90b} associated the weakness of NH$_3$(1,1) and
(2,2) toward the continuum source either with photoionization or high
temperatures present in the nuclear region. Because of the detection
of NH$_3$(6,6), which is at 412~K above ground, the second scenario is
indeed the most plausible explanation. More recent results in
NH$_3$(1,1) and (2,2) are consistent with the earlier results
(M. Lebr{\'o}n, private communication).

Below, we discuss the three NH$_3$(6,6) peaks in turn.

\subsection{(6,6) peak}
When comparing the NH$_3$(6,6) map with the positions of the GMCs, we
notice that the {\it (6,6) peak} is remarkably close to GMC C (figure
\ref{positions.fig}). The two positions are in fact coincident to
within the absolute positional uncertainties of the NH$_3$ map.  We
conclude that the {\it (6,6) peak} and GMC C are likely the same
structure. Previous studies have demonstrated that the strongest star
formation areas in the central region of IC~342 are GMCs B and C,
which are both located where the arms meet the ring. Therefore,
NH$_3$(6,6) traces areas where the star formation activity has been
enhanced.

\begin{figure}
\begin{center}
\epsscale{1}
\plotone{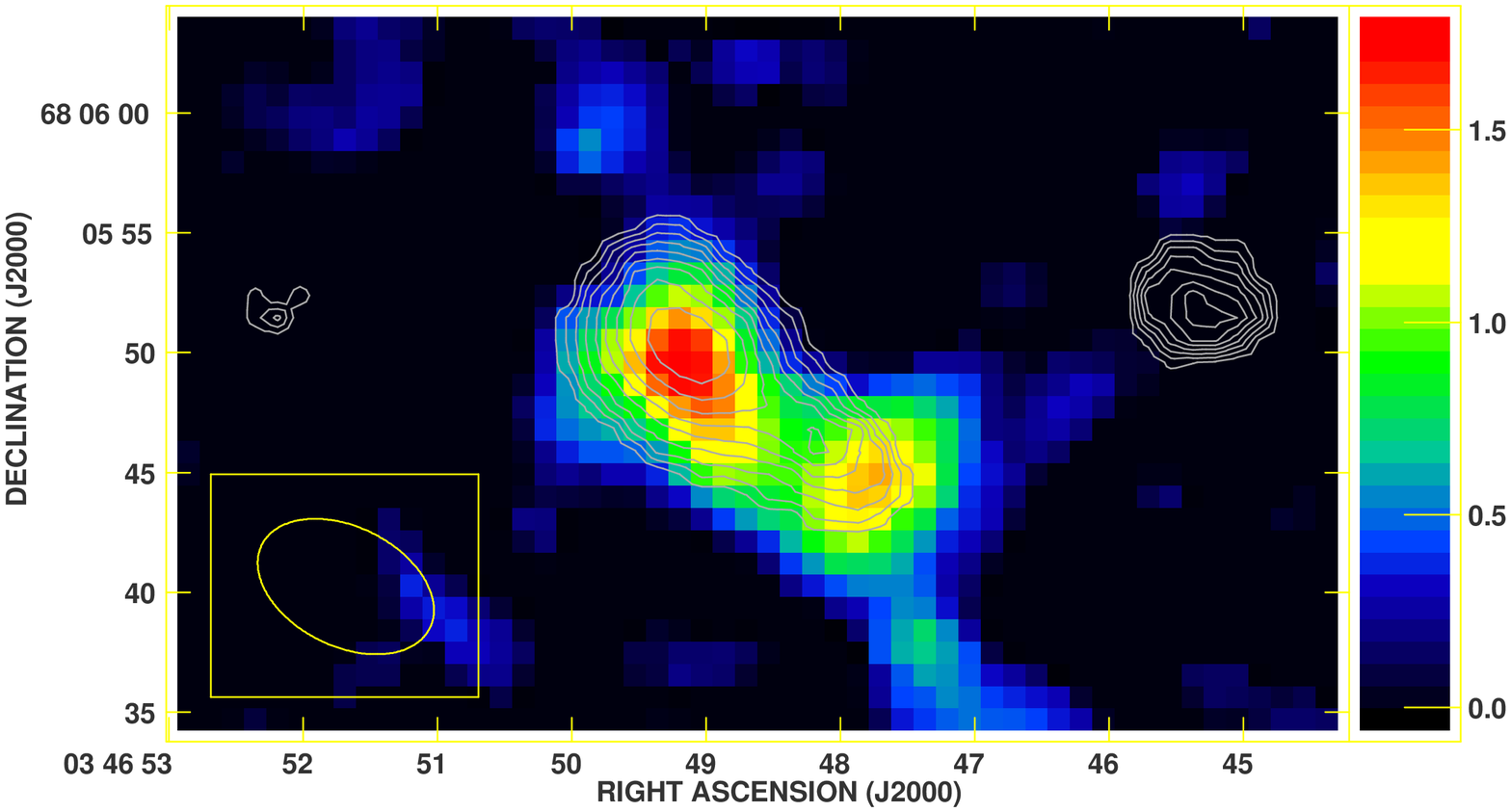}
\end{center}
\caption{NH$_3$(6,6) integrated intensity in contours. HC$_3$N(10-9) integrated intensity in false-color scale. Contour levels are as in figure 1. The false-color scale is in mJy~Beam$^{-1}$~km~s$^{-1}$ \citep{mei05}. \label{hc3n.fig}}
\end{figure}

The HC$_3$N(10-9) emission from \citet{mei05} has a structure similar
to that of the NH$_3$(6,6) emission (figure \ref{hc3n.fig}). In both
maps, the strongest line emission is found at the position of the {\it
(6,6) peak}. We compare the line profiles detected in HC$_3$N(10-9) by
\citet{mei05} with that in NH$_3$(6,6) at the position of the GMC that
is closer to our detected peak (GMC C) and we find that they have very
similar velocities. Thus HC$_3$N(10-9) and NH$_3$(6,6) appear to trace the
same high-density, high-temperature material.

When comparing the NH$_3$(6,6) results with those from the HNC(1-0) map by
\citet{mei05} we find that the emission of the GMC close to the
position of the {\it (6,6) peak} (GMC C) is weaker than the emission
arising from another region of the map (figure 5). The different behavior might indicate that HNC(1-0) is a density tracer while NH$_3$(6,6) is a better temperature tracer. In fact, \citet{mei05} note that HNC is more strongly detected in areas of high volume density, not only high column density. 

The distribution of N$_2$H$^+$(1-0) is very similar to that of HNC(1-0) (figure
\ref{n2h+.fig}), although the N$_2$H$^+$(1-0) emission is slightly more
extended towards the north, tracing the northern arm of the
mini-spiral. The N$_2$H$^+$(1-0) emission peaks at the location of GMC N,
which is not close to the {\it (6,6) peak} (see figure 1).  N$_2$H$^+$
is believed to trace dense quiescent gas \citep{wom92}. The lack of
coincidence between the N$_2$H$^+$(1-0) and the NH$_3$(6,6) maps indicates
that the ion is tracing the areas where the star formation seems to be
less intense or non-existent, while the NH$_3$(6,6) is present in active
star formation regions.

The {\it (6,6) peak} coincides with the CO(2-1) peak measured by
\citet{schi03} (figure \ref{co21.fig}). The CO(2-1) traces the eastern
and western parts of the two molecular spiral arms where they form the
molecular ring. The NH$_3$(6,6) appears to extend outside of the
eastern arm, but the spatial coincidence is excellent when the
different angular resolutions of the two data-sets are
considered. This coincidence of NH$_3$(6,6) and CO(2-1) suggests that
the gaseous material flowing inwards along the northern arm of the
``mini-spiral'' is interacting with the inner ring, heating GMC C, and
therefore producing the CO(2-1) emission.  This interaction may also
be triggering a burst of star formation at the same location.

Like in the CO(2-1) map, the NH$_3$ {\it (6,6) peak} appears to be
offset by 1'' to the SE as compared to the CO(1-0) map by
\citet{hel03} (figure \ref{co10.fig}). In this case, the angular
resolutions of the two maps are similar. The slight offset may not be
significant given the absolute positional error of about 1''. The
CO(1-0) traces the more extended and colder gas that forms the
S-shaped bar, whereas the NH$_3$(6,6) is more compact and detects the
warmer material accumulated in the regions where the inflow from the
bar meets the inner ring. However, the NH$_3$(6,6) and the CO(1-0)
maps do seem to be well-correlated, since both main peaks are
detected.

When comparing the NH$_3$(6,6) map with the 6~cm continuum map, we
find that the {\it (6,6) peak} appears at the edge of the 6~cm
continuum emission (figure \ref{6cm.fig}). The 6~cm emission seems to
have a ``tail'' that ends at the position of the {\it (6,6) peak}.
However, there is no evidence for a diffuse continuum component in
this area. \citet{bec80} and \citet{tur83} found that the 6~cm
continuum map of the center of IC~342 was mainly dominated by
free-free thermal emission.  However, in the northern part of the CND
where the {\it (6,6) peak} is located, the emission was mostly from
synchrotron origin.  Thus, there is synchrotron emission associated
with the {\it (6,6) peak}, but not with its surrondings.  A possible
explanation for these observations could be that the {\it (6,6) peak}
is located in the position of a very concentrated burst of star
formation that has developed some supernovae, but no such activity has
taken place outside of this position.

In summary, the presence of a strong star formation region where the {\it (6,6) peak} is located, and a comparison of the NH$_3$(6,6) distribution with other molecular lines, suggest that the {\it (6,6) peak} has the characteristics of a column density and temperature peak.

\subsection{Continuum peak}
The NH$_3$(6,6) {\it continuum peak} is offset by 2''-3'' from GMCs B,
A and N, which trace an arc around it (figure
\ref{positions.fig}). This offset appears significant, especially
since the NH$_3$(6,6) {\it continuum peak} coincides well with the
6~cm continuum peak (figure \ref{6cm.fig}). When comparing GMC B and
the {\it continuum peak} spectra, we notice that even though the {\it
continuum peak} spectrum is stronger, the integrated intensity is very
similar. As remarked before, GMC B is a very strong star formation
area. On the other hand, no trace of star formation has been detected
towards GMC~A to date.  The lack of a strong NH$_3$(6,6) detection at
the position of GMC~A supports the theory that there is no current
star formation in this region.  GMC A has a NH$_3$(6,6) spectrum only
slightly more intense than the background emission-free spots (figure
\ref{spectra.fig}). We do not find a differentiated feature in the
integrated intensity map where GMC N should be. However, the spectrum
at this position is clearly above noise-level.

The HC$_3$N(10-9) map from \citet{mei05} shows a peak near the {\it
continuum peak} (figure \ref{hc3n.fig}). As in the NH$_3$(6,6) map,
the {\it continuum peak} seems weaker than the {\it (6,6) peak}. Since
GMC B, which seems to be the GMC closer to the {\it continuum peak},
is the warmest region of the nucleus of IC~342, the fact that
NH$_3$(6,6) and HC$_3$N(10-9) are weaker at this point may suggest
that the total column density is not a maximum at this position.

\begin{figure}
\begin{center}
\epsscale{1}
\plotone{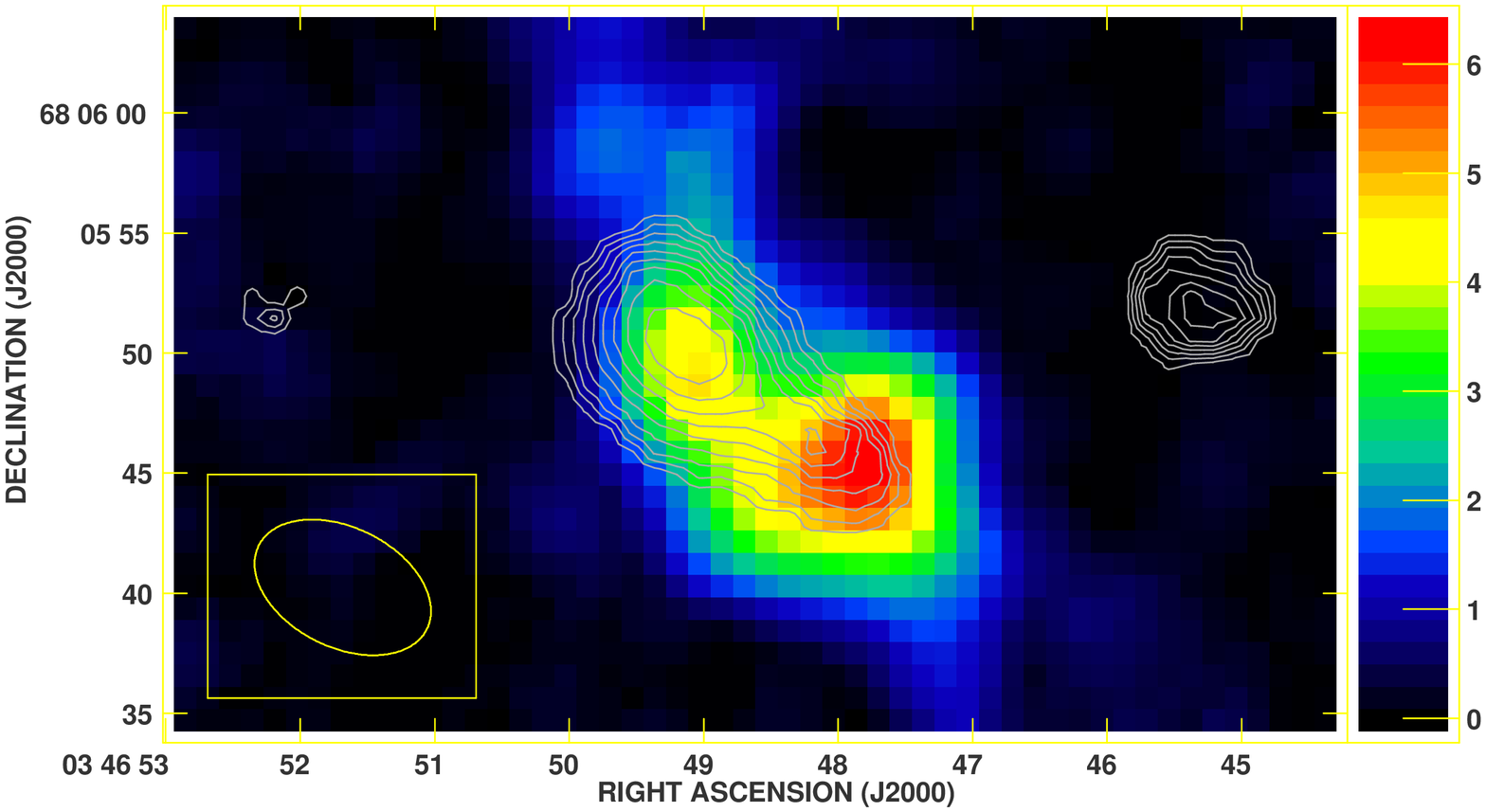}
\end{center}
\caption{NH$_3$(6,6) integrated intensity in contours. HNC(1-0) integrated intensity in false-color scale. Contour levels are as in figure 1. The false-color scale is in mJy~Beam$^{-1}$~km~s$^{-1}$ \citep{mei05}. \label{hnc.fig}}
\end{figure}

\begin{figure}
\begin{center}
\epsscale{1}
\plotone{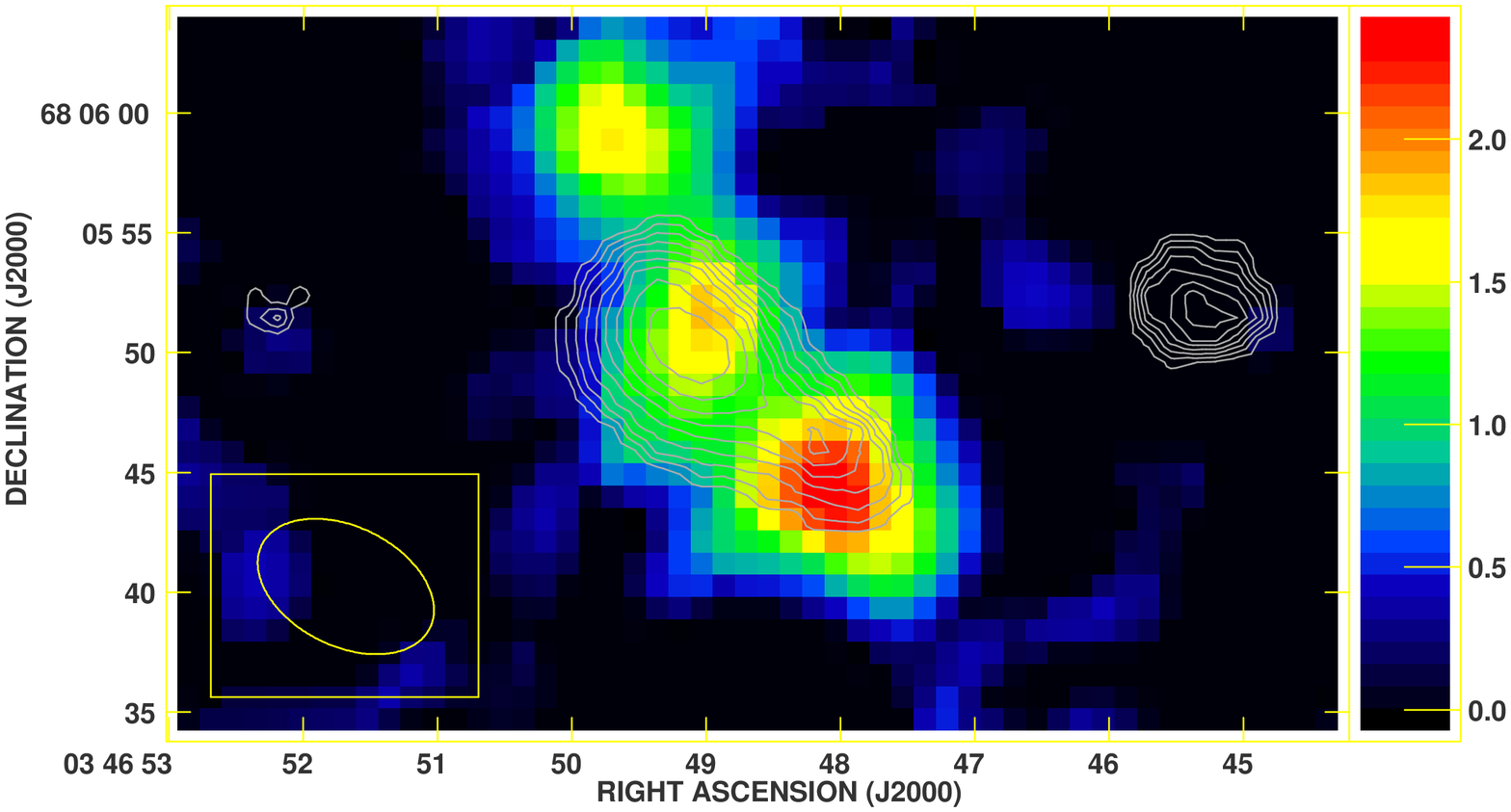}
\end{center}
\caption{NH$_3$(6,6) integrated intensity in contours. N$_2$H$^+$(1-0) integrated intensity in false-color scale. Contour levels are as in figure 1. The false-color scale is in mJy~Beam$^{-1}$~km~s$^{-1}$ \citep{mei05}. \label{n2h+.fig}}
\end{figure}

\begin{figure}
\begin{center}
\epsscale{1}
\plotone{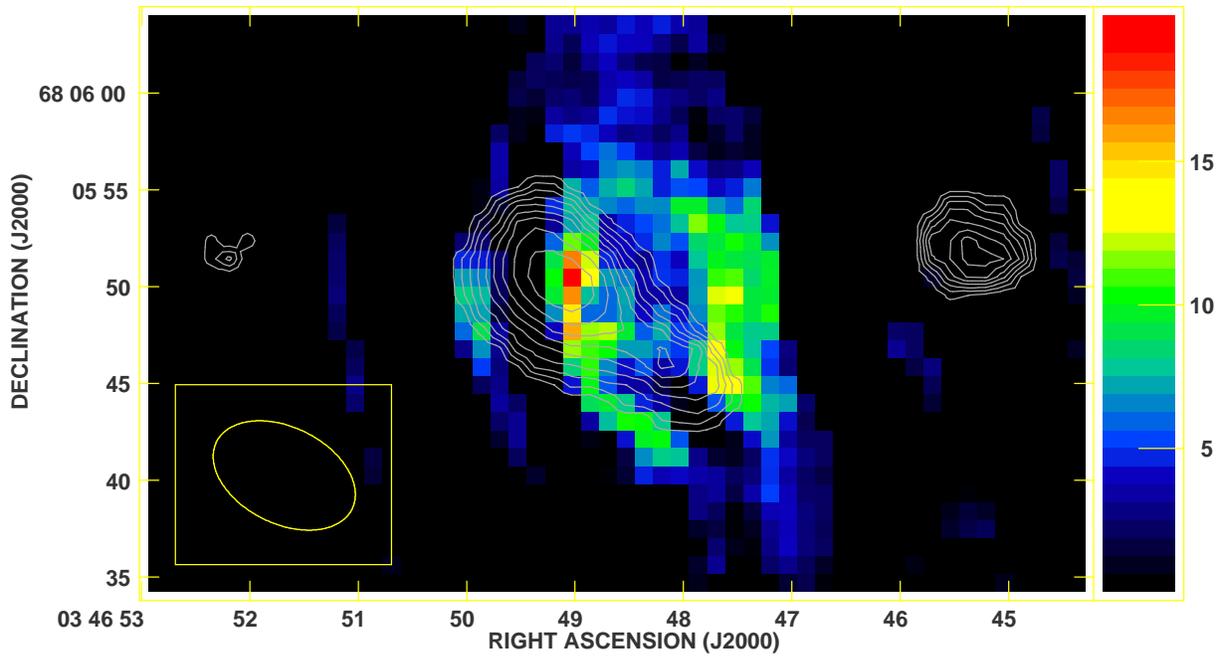}
\end{center}
\caption{NH$_3$(6,6) integrated intensity in contours. CO(2-1) integrated intensity is in false-color scale and traces the eastern and western parts of the molecular ring. Contour levels are as in figure 1. The false-color scale is in mJy~Beam$^{-1}$~km~s$^{-1}$ \citep{schi03}. \label{co21.fig}}
\end{figure}

\begin{figure}
\begin{center}
\epsscale{1}
\plotone{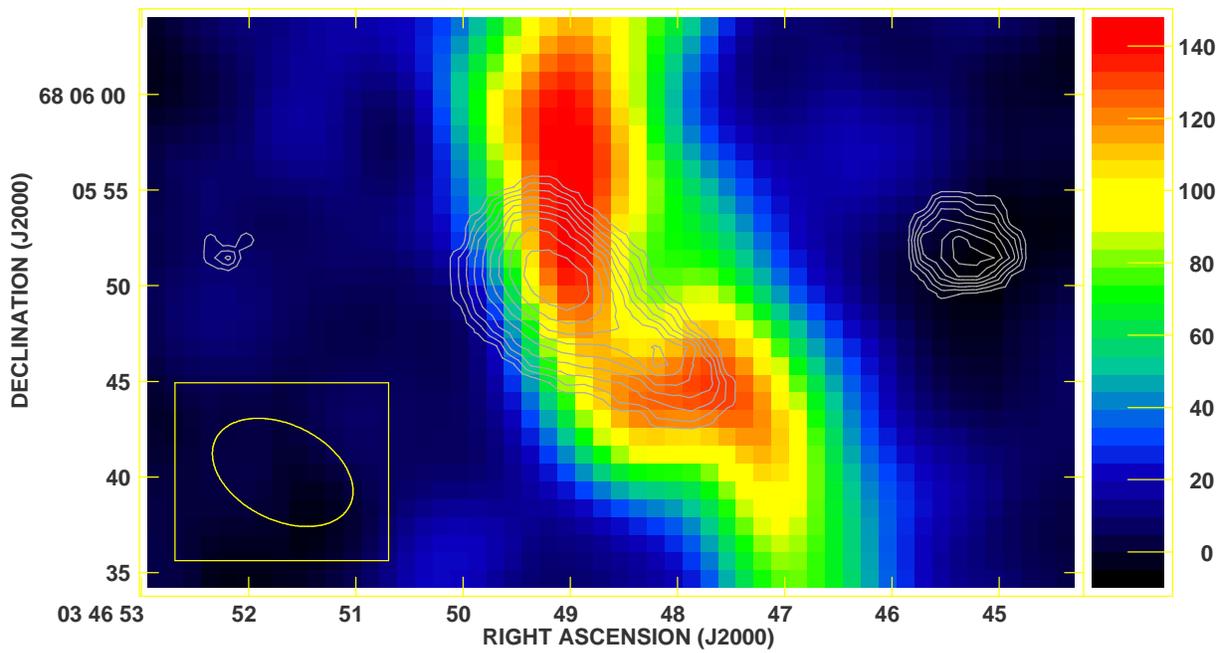}
\end{center}
\caption{NH$_3$(6,6) integrated intensity in contours. CO(1-0) integrated intensity is shown in false-color scale, and traces the mini-spiral. Contour levels are as in figure 1. The false-color scale is in mJy~Beam$^{-1}$~km~s$^{-1}$ \citep{hel03}. \label{co10.fig}}
\end{figure}

\begin{figure}
\begin{center}
\epsscale{1}
\plotone{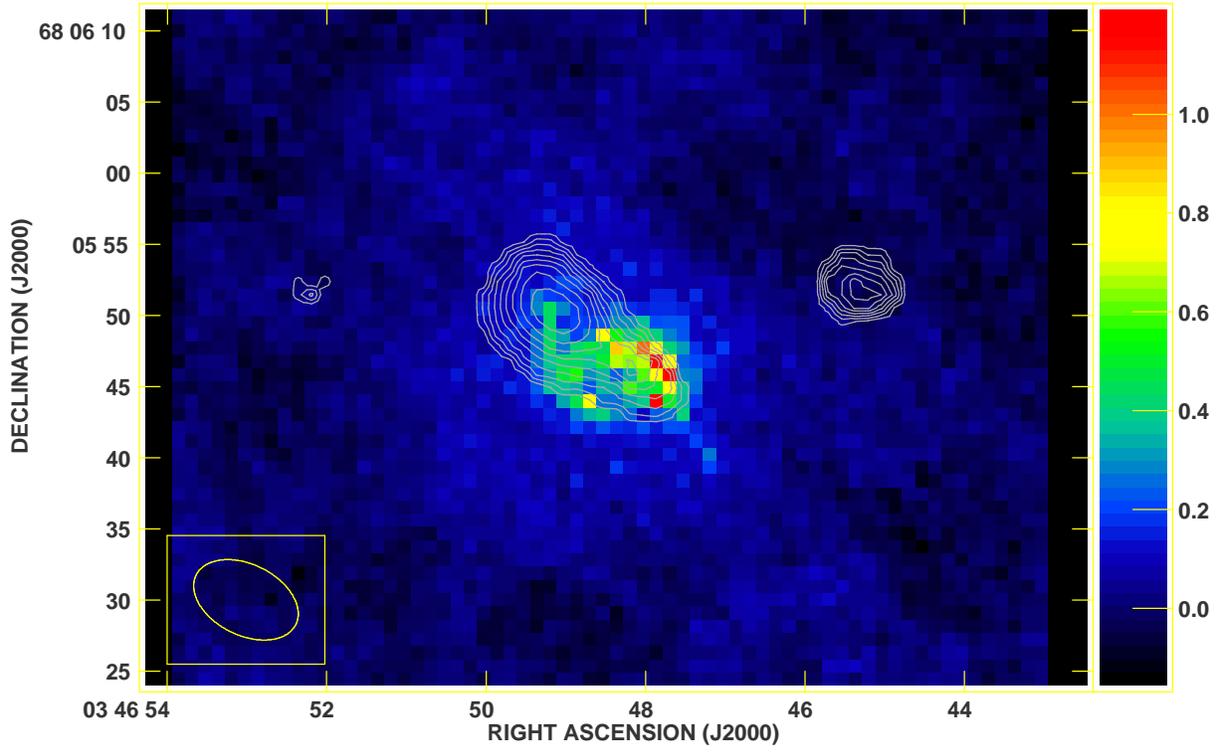}
\end{center}
\caption{NH$_3$(6,6) integrated intensity in contours. VLA 6~cm continuum integrated intensity in false-color scale. Contour levels are as in figure 1. The false-color scale is in mJy~Beam$^{-1}$.\label{6cm.fig}}
\end{figure}

\begin{figure}
\begin{center}
\epsscale{1}
\plotone{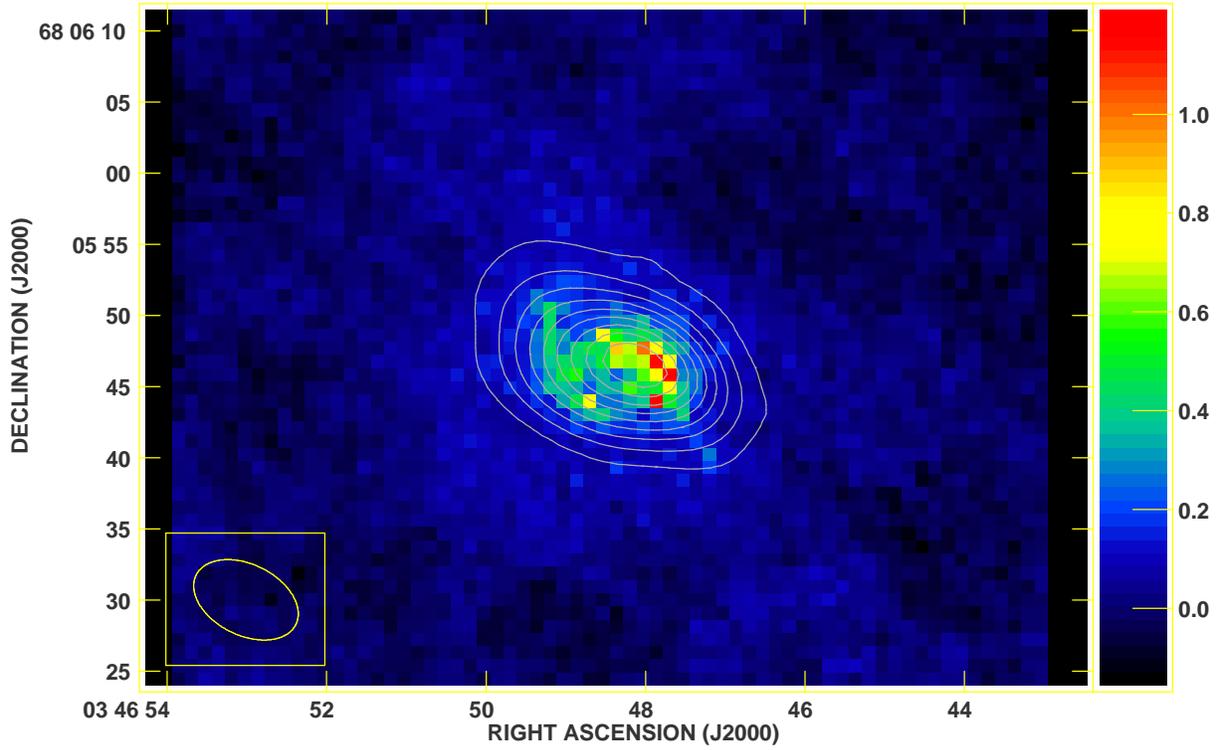} 
\end{center}
\caption{1.3~cm continuum integrated intensity in contours. VLA 6~cm continuum integrated intensity in false-color scale. Contour levels are in steps of 10\% of the continuum intensity peak, from 13~mJy~Beam$^{-1}$ to 114~mJy~Beam$^{-1}$. The false-color scale is also in mJy~Beam$^{-1}$.\label{cont_cont.fig}}
\end{figure}

The {\it continuum peak} is offset by 2'' north of the secondary peak
in HC$_3$N(10-9) and the primary peak detected in HNC(1-0) by
\citet{mei05} (figure \ref{hnc.fig}). This offset appears to be
significant.  A similar offset between NH$_3$(6,6) and N$_2$H$^+$(1-0) is
also observed.  These results suggest that neither HC$_3$N(10-9) nor
HNC(1-0) nor N$_2$H$^+$(1-0) are tracing a temperature peak, which is in
fact better traced by NH$_3$(6,6).

The {\it continuum peak} is a very weak feature in the CO(2-1) map
\citep{schi03}. Unlike NH$_3$(6,6), CO(2-1) does not detect the gas
associated with the heating source located at the {\it continuum
peak}, but instead traces the gas surrounding it (figure
\ref{co21.fig}).

The {\it continuum peak} also appears offset by 3'' from the secondary
peak detected in the CO(1-0) map by \citet{hel03} (figure
\ref{co10.fig}). As in the NH$_3$(6,6) map, this secondary peak is
weaker than the one that coincides with the {\it (6,6) peak}, and
although the kind of gas traced by the two molecules have different
temperatures, the overall structure has some similarities.

Finally, as expected, the {\it continuum peak} is located close to the
6~cm continuum emission peak (figure \ref{6cm.fig}). The 6~cm
continuum emission seems to trace an arc structure around the
NH$_3$(6,6) {\it continuum peak} rather than being concentrated at
that same spot.  In this case, the angular resolution is different by
a factor of 10. With higher resolution, the NH$_3$(6,6) emission could
be even better correlated with the 6~cm continuum structure.  The good
correspondence with the radio continuum peak, and the offset of a few
arcseconds relative to other molecular peaks which sample lower
temperature gas, suggests that the NH$_3$(6,6) emission could be
tracing a hot molecular component near the nucleus. \citet{isr03}
suggest that the nucleus of IC~342 has two different molecular mass
components, a cold and dense component, which represents about one
third of the total molecular mass, and a less dense and hotter
component. A similar situation has also been observed in the center of
the Milky Way. By looking at six NH$_3$ metatestable inversion
transitions (from (1,1) to (6,6)) in the Galactic Center region
(575~pc~$\times$~145~pc), \citet{hut93} find the distribution of the
transition lines to differ from one another. While lower transition
lines look similar, the higher transitions often look very
different. In fact, \citet{hut93} find that molecular gas in the
nuclear region of the Galaxy is best modeled as a mixture of hot
($\sim$~200~K) and cold (25~K) gas, in which 75$\%$ of the gas is in
the cold component. This ``two-temperature'' structure continues to
hold even within 5~pc of the nucleus \citep{her05}. Furthermore, studies by
\citet{her02}, with a similar velocity resolution to ours, find NH$_3$(6,6) to be clearly concentrated in the nucleus of the Milky Way. Obviously, the Galactic Center study refers to a much more concentrated and less massive material than the material detected in IC~342, given the different size-scales (2.7~kpc~$\times$~4.8~kpc for IC~342, 10~pc~$\times$~10~pc for the Galactic Center). Therefore, the mechanism that drives the NH$_3$(6,6) in the nuclear region of IC~342, which is likely to be a star formation process, might be more extensive spatially than the mechanism that is working in the Galactic Center. Nevertheless, given the similar behaviour we find in the nuclear regions of IC~342 and the Galactic Center, this suggests that the {\it continuum peak} could be tracing the hot component present in the nucleus of IC~342, i.e. the {\it continuum peak} is a temperature peak.

\subsection{West peak}
The {\it west peak} has no counterpart in any other molecular line map
that has been published to date. However, the careful reduction
process that we have used gives us confidence to conclude that this
structure is not a noise feature. In addition, the lack of a symmetric
counterpart on the other side of the nucleus of IC~342 leads us to
conclude that the {\it west peak} is unlikely a sidelobe effect.  The
{\it west peak} is not very strong, but it is a remarkable feature
because it is detected over a wide velocity range. We compare the {\it
west peak} spectrum to three randomly chosen noise-level spectra in
figure \ref{spectra.fig}.  We find that, although weak, the overall flux level associated with the {\it west peak} is above the noise level.

The comparison of the NH$_3$(6,6) emission map with the latest VLA 6~cm continuum emission map from Ch.-W. Tsai (A and C configurations combined) shows no evidence of radiocontinuum emission at this location (figure \ref{6cm.fig}). 

%It is very remarkable that the {\it west peak} appears centered on a mostly ``empty'' region of the spiral structure traced by the 6~cm continuum emission (there is some faint 6~cm emission at this position) (figure \ref{6cm.fig}). There are large concentrations of emission just NW, NE and SW of this feature (apart from the nucleus) but the {\it west peak} is mainly isolated. As the 6~cm emission likely traces synchrotron emission, any star formation activities at this position may be still too young to produce supernovae and associated synchrotron emission. The location of the {\it west peak} may be where the 6~cm spiral arm terminates at a nuclear bar or a resonance point.

The presence of NH$_3$(6,6) at positions where there have been no
previous molecular detections has occurred elsewhere. In the Milky
Way, NH$_3$(6,6) has been detected within 2~pc of the Galactic Center
with no corresponding emission from other molecular tracers
\citep{her02}. This molecular emission (termed the
``high-line-ratio-cloud'') is thought to be the result of absorption
of low-energy transitions (in particular NH$_3$(1,1), (2,2) and (3,3))
by cool material along the line-of-sight \citep{her05}. Though the
size scale of the Galactic Center emission ($<$~2~pc) is much smaller
than the size scales traced in our IC~342 data, it is possible that a
similar geometrical effect could explain the lack of low-energy
molecular emission associated with the IC~342 {\it west peak}.

\subsection{Mass estimates}

Since we have detected three different peaks, we separate the
NH$_3$(6,6) structure into three different areas and calculate the
mass in each.

Because we have only measured one transition line, NH$_3$(6,6), we
cannot directly calculate the excitation temperature of the gas.
Theoretically, the NH$_3$(6,6) opacity could be measured using the
hyperfine splitting of the NH$_3$(6,6) emission line.  However, due to
the weakness of the satellite lines (roughly 3\% of the main hyperfine
line for optically thin gas) and the large linewidths, this
calculation is not possible for our data. We therefore assume
optically thin emission ($\tau$~$\ll$~1) and also consider the source
to fill the clean beam giving
\begin{equation}
{\it T_{ex} \tau \cong T_{b}}~.
\end{equation}

We can obtain the T$_b$ directly from the line profile, using
\begin{equation}
{\it T_b~=~\frac {\lambda^2}{2 k \Omega} S_{\nu}}~,
\end{equation}
where $\lambda$ is the wavelength, k is Boltzmann's constant, $\Omega$ the solid angle and S$_\nu$ is the peak flux density.

The column density is calculated using
\begin{equation}
{\it N_{JK}~=~4.1 \times 10^{-20} cm^{-2} \frac {\tau \nu T_{ex} \Delta v}{A_{10} k_{JK}}}~.
\end{equation}
For NH$_3$(6,6), k$_{66}$~=~0.969, A$_{10}$~=~3.38~$\times$~10$^{-7}$~s$^{-1}$, and $\nu$~=~25.056~GHz \citep{her03}. 

Once we know the value of the column density, we obtain the molecular mass through the following equation:
\begin{equation}
{\it M_{H_2}~=~N_{66}~\times~Area~\times~\frac {1}{f_{66}}~\times~\left[\frac {N(H_2)}{N(NH_3)}\right]~\times~m_{H_2}}~,
\end{equation}
where f$_{66}$ is the fraction of NH$_3$ molecules in the (6,6) state,
and is on the order of 0.1 for temperatures higher than 150~K. We use
$\frac {N(H_2)}{N(NH_3)}$~=~10$^{7.5}$ from \citet{mau03}, which is
the relative abundance of H$_2$ to NH$_3$ in fairly warm environments
(T~$>$~150~K).  

The resulting mass estimates for each NH$_3$(6,6) peak are given in table 1.

We detect a total warm gas mass of
8.2~$\times$~10$^{6}$~M$_\odot$, clearly much larger than the amount of warm gas mass detected in the interior of the Galactic Center Circumnuclear Disk by \citet{her02}, 10$^{4}$~M$_\odot$ (inner 1.5~pc around Sgr~A*). \citet{rig02} reported a warm mass of 5.0~$\times$~10$^{6}$~M$_\odot$ measuring the quadrupole transition lines of H$_2$ using the Infrared Space Observatory (ISO), assuming a distance of 3.6~Mpc and an area coverage of 14''~$\times$~27''. If we assume the same distance and coverage (the {\it west peak} would be left outside), the warm gas mass detected would be 9.4~$\times$~10$^{6}$~M$_\odot$. \citet{rig02} estimate that the 2.5\% of the gas in IC~342 is warm, using previous published data from \citet{aal95} in various $^{12}$CO, $^{13}$CO and C$^{18}$O transitions for the total molecular gas mass value, measured with a variety of telescopes. Using the same value for the total molecular gas mass (and correcting for the distance) we find that the percentage increases to 4.7\%. Additionally, \citet{mau03} report the 10\% of the total molecular gas in IC~342 to be warm. \citet{mau03} use the warm gas mass value reported by \citet{rig02} corrected for the assumed distance (1.8~Mpc), and compare it to the total molecular gas mass measured using $^{13}$C$^{16}$O data taken with the Heinrich-Hertz-Telescope. Assuming the same distance (1.8~Mpc) and amount of total molecular gas mass than \citet{mau03}, and considering the three peaks, since the beam of Heinrich-Hertz-Telescope (34'') covers all three of them, the amount of warm gas mass would be 7.2~$\times$~10$^{5}$~M$_\odot$, which represents the 5.5\% of the total molecular gas mass. In summary, using two different values for the total molecular gas mass (from \citet{rig02} and \citet{mau03}), and two different area coverages, we still find that the percentage of warm gas mass to total molecular gas mass is $\sim$ ~5\%.

%%%%%%%%%%%%%%%%%%%%%%%%%%%%%%%% SUMMARY %%%%%%%%%%%%%%%%%%%%%%%%%%%%%%%%

\section{Summary}

We have imaged the nuclear region of IC~342 in NH$_3$(6,6) using the
VLA.  Three emission peaks are detected. The {\it (6,6) peak}
corresponds well with GMC C and is probably a column density and
temperature peak driven by a burst of star formation at the position
where inflowing gas accumulates at a nuclear ring. The {\it continuum
peak} is offset from GMCs B, N and A, but coincides with the 1.3~cm
continuum peak. The 6~cm continuum peak is also very close, and due to
the different resolution betweeen the two maps, we can not rule out
that the two peaks could be in fact coincident. The {\it continuum
peak} is likely a temperature peak corresponding to very hot gas
similar to the hot gas seen in our own Galactic Center around
Sgr~A*. The {\it west peak} is a tentative detection with not known molecular or continuum emission counterpart. The fact that the {\it west
peak} remains undetected in lower transition lines might be due to
absorption along the line-of-sight, as is the case for the
``high-line-ratio-cloud'' in the Galactic Center.  The presence of GMC
N, as detected by \citet{mei05}, could not be positively confirmed in
NH$_3$(6,6) emission. NH$_3$(6,6) emission above noise-level is detected at
the position where this GMC has been reported. However, we did not
detect a distinctive feature at this location.

%%%%%%%%%%%%%%%%%%%%%%%%%%%%%% ACKNOWLEDGMENTS %%%%%%%%%%%%%%%%%%%%%%%%%%%%%%

\acknowledgments

We thank C.-W. Tsai and M. Lebr{\'o}n for sharing their results in
advance of publication. Thanks are also due to E. Schinnerer and
D. Meier for providing the maps we have used for comparison and to
T. Helfer for kindly answering all our questions regarding her
results.  During the development of this study, MM-C has been
supported by an Academia Sinica Institute of Astronomy and
Astrophysics (ASIAA) fellowship and a Smithsonian Institution Visiting
Student Grant.

%%%%%%%%%%%%%%%%%%%%%%%%%%%%%%% BIBLIOGRAPHY %%%%%%%%%%%%%%%%%%%%%%%%%%%%%%%%

%%%%%%%%%%%%%%%%%%%%%%%%%%%%%%%%%%%%%%%%%%%%%%%%%%%%%%%%%%%%%%%%%%%%%%%%%%%%%

%%%%%%%%%%%%%%%%%%%%%%%%%%%%% FIGURES %%%%%%%%%%%%%%%%%%%%%%%%%%%%%%%%%%%%%%%

%%%Continuum + Line with 3 peaks

%%%Spectral Points

%%%Spectra

%%%Line + HC3N

%%%Line + HNC

%%%Line + N2H+

%%%Line + CO(2-1)

%%%Line +CO(1-0)

%%%Line + 6cm

%%%Cont + 6cm

%%%%%%%%%%%%%%%%%%%%%%%%%%%%%%TABLE%%%%%%%%%%%%%%%%%%%%%%%%%%%%%%%%%%%%%%%%%%
\clearpage

\begin{table}
\begin{center}
\caption{Calculated warm masses for the three detected peaks for a distance of 3.3~Mpc.}
\begin{tabular}{l|l|l|l|l|l|l} \tableline \tableline
Peak & Line flux & T$_b$ & $\Delta$v & N$_{66}$ & Area  & M$_{H_2}$  \\  
  & 10$^{-3}$~Jy~Beam$^{-1}$ & K & km~s$^{-1}$ & 10$^{13}$~cm$^{-2}$ & 10$^{4}$~pc$^{2}$ & 10$^{6}$~M$_\odot$  \\ \tableline
{\it (6,6) peak} & 2.1 & 0.12 & 40  & 3.7 & 2.6  & 4.8 \\ \tableline
{\it Continuum peak} & 1.4 & 0.08 & 50  & 3.1 & 1.2 & 1.8 \\ \tableline
{\it West peak} & 1.1 & 0.06 & 80  & 4.0 & 0.8 & 1.6  \\ \tableline  
\end{tabular}
\end{center}
\end{table}

%%%%%%%%%%%%%%%%%%%%%%%%%%%%%%%%%%%%%%%%%%%%%%%%%%%%%%%%%%%%%%%%%%%%%%%%%%%%%


\begin{thebibliography}{}

\bibitem[Aalto et al.(1995)]{aal95} Aalto, S., Booth, R.S., Black, J.H. \& Johansson, L.E.B. 1995, \aap, 300, 369
\bibitem[Becklin et al.(1980)]{bec80} Becklin, E.E., Gatley, I., Mathews, K., Neugebauer, G. Sellgren, K., Werner, M.W. \& Wynn-Williams, C.G. 1980, \apj, 236, 441 
\bibitem[Beuther et al.(2005)] {beu05} Beuther, H., Thorwirth, S., Zhang, Q., Hunter, T.R., Megeath, S.T., Walsh, A.J. \& Menten, K.M. 2005, \apj, 627, 834
\bibitem[B{\"o}ker et al.(1997)]{bok97} B{\"o}ker, T., F{\"o}ster-Schreiber, N.M. \& Genzel, R. 1997, \aj, 114, 1883 
\bibitem[B{\"o}ker et al.(1999)]{bok99} B{\"o}ker, T., van der Marel, R.P. \& Vacca, W.D. 1999, \apj, 118, 831
\bibitem[Downes et al.(1992)]{dow92} Downes, D., Radford, S.J.E., Guilloteau, S., Gu{\'e}lin, M., Greve, A. \& Morris, D. 
1992, \aap, 262, 424    
\bibitem[Helfer \& Blitz(1993)]{hel93} Helfer, T.T. \& Blitz, L. 1993, \apj, 419, 86 
\bibitem[Helfer et al.(2003)]{hel03} Helfer, T.T., Thornley, M.D., Regan, M.W., Wong, T., Sheth, K., Vogel, S.N., Blitz, L. \& Bock, D.C.-J. 2003, \apjs, 145, 259 
\bibitem[Henkel et al.(2000)]{hen00} Henkel, C., Mauersberger, R., Peck, A.B., Falcke, H. \& Hagiwara, Y. 2000, \aap, 361, L45
\bibitem[Herrnstein \& Ho(2002)]{her02} Herrnstein, R.M. \& Ho, P.T.P. 2002, \apj, 579, L83 
\bibitem[Herrnstein(2003)]{her03} Herrnstein, R.M. 2003, Ph.D Thesis, Harvard University
\bibitem[Herrnstein \& Ho(2005)]{her05} Herrnstein, R.M. \& Ho, P.T.P. 2005, \apj, 620, 287 
\bibitem[Ho, Martin, \& Ruf(1982)]{ho82} Ho, P. T. P., Martin, R. N., \& Ruf, K. 1982, \aap, 113, 155
\bibitem[Ho \& Martin(1983)]{ho83} Ho, P.T.P. \& Martin, R.N. 1983, \apj, 272, 484
\bibitem[Ho et al.(1987)]{ho87} Ho, P.T.P., Turner, J.L. \& Martin, R.N. 1987, \apj, 322, L67 
\bibitem[Ho et al.(1990)]{ho90b} Ho, P.T.P., Martin, R.N., Turner, J.L. \& Jackson, J.M. 1990, \apj, 355, L19
\bibitem[H{\"u}ttemeister et al.(1993)]{hut93} H{\"u}ttemeister, S., Wilson, T.L., Bania, T.M. \& Mart{\'{\i}}n-Pintado, J. 1993, \aap, 280, 255
\bibitem[H{\"u}ttemeister et al.(1995)]{hut95} H{\"u}ttemeister, S., Henkel, C., Mauersberger, R., Brouillet, N., Wiklind, 
T. \& Millar, T. J. 1995, \aap, 295, 571 
\bibitem[Ishizuki et al.(1990)]{ish90} Ishizuki, S., Kawabe, R., Ishiguro, M., Okumura, S.K., Morita, K.I., Chikada, Y.\& 
Kasuga, T. 1990, \nat, 344, 224 
\bibitem[Israel \& Baas(2003)]{isr03} Israel, F.P. \& Baas, F. 2003, \aap, 404, 495 
\bibitem[Jackson et al.(1996)]{jac96} Jackson, J.M., Heyer, M., Paglione, T.A.D. \& Bolatto, A.D. 1996, \apj, 456, L91 
\bibitem[Karachentsev(2005)]{kar05} Karachentsev, I.D. 2005, \apj, 129, 178
\bibitem[Lo et al.(1984)]{lo84} Lo, K.Y., Berge, G.L., Claussen, M.J., Heiligman, G.M., Leighton, R.B., Masson, C.R., Moffet, A.T., Phillips, T.G., Sargent, A.I., Scott, S.L., Wannier, P.G. \& Woody, D.P. 1984, \apj, 282, L59 
\bibitem[Martin \& Ho(1979)]{mar79} Martin, R.N. \& Ho, P.T.P. 1979, \aap, 74, L7
\bibitem[Martin \& Ho(1986)]{mar86} Martin, R.N. \& Ho, P.T.P. 1986, \apj, 308, L7
\bibitem[Mauersberger et al.(2003)]{mau03} Mauersberger, R., Henkel, C., Wei\ss, A., Peck, A.B. \& Hagiwara, Y. 2003, \aap, 403, 561
\bibitem[Meier et al.(2000)]{mei00} Meier, D.S., Turner, J.L. \& Hurt, R.L. 2000, \apj, 531, 200 
\bibitem[Meier \& Turner(2001)]{mei01} Meier, D.S. \& Turner, J.L. 2001, \apj, 551, 687 
\bibitem[Meier \& Turner(2005)]{mei05} Meier, D.S. \& Turner, J.L. 2005, \apj, 618, 259
\bibitem[Ott et al.(2005)] {ott05} Ott, J., Weiss, A., Henkel, C. \&  Walter, F. 2005, \apj, 629, 767
\bibitem[Rigopoulou et al.(2002)]{rig02} Rigopoulou, D., Kunze, D., Lutz, D., Genzel, R. \& Moorwood, A.F.M. 2002, \aap, 389, 374  
\bibitem[Saha et al.(2002)]{sah02} Saha, A., Claver, J. \& Hoessel, J.G. 2002, \aj, 124, 839
\bibitem[Schinnerer et al.(2003)]{schi03} Schinnerer, E., B{\"o}ker, T. \& Meier, D.S. 2003, \apj, 591, L115 
\bibitem[Schulz et al.(2001)]{schu01} Schulz, A., G{\"u}rsten, R., K{\"o}ster, B. \& Krause, D. 2001, \aap, 371, 25 
\bibitem[Takano et al.(2002)]{tak02} Takano, S., Nakai, N. \& Kawaguchi, K. 2002, \pasj, 54, 195 
\bibitem[Turner et al.(1993)]{tur93}Turner, J.L., Hurt, R.L. \& Hudson, D.Y. 1993, \apj, 413, L19
\bibitem[Turner \& Hurt(1992)]{tur92} Turner, J. L., \& Hurt, R. L. 1992, \apj, 384, 72 
\bibitem[Turner \& Ho(1983)]{tur83} Turner, J.L. \& Ho, P.T.P. 1983, \apj, 268, L79
\bibitem[Wei$\ss$ et al.(2001)]{wei01} Wei\ss, A., Neininger, N., Henkel, C., Stutzki, J. \& Klein, U. 2001, \apj, 554, L143
 
\bibitem[Womack et al.(1992)]{wom92} Womack, M. Ziurys, L.M. \& Wyckoff, S. 1992, \apj, 393, 188
\end{thebibliography}
\end{document}